\documentclass[journal]{IEEEtran}
\usepackage{xcolor}
\usepackage{tabularx,booktabs,siunitx}
\usepackage{float}
\usepackage{amsmath}
\usepackage{amsthm}
\usepackage{graphicx}
\usepackage{graphics}
\usepackage{siunitx}
\usepackage{multicol}
\usepackage{stfloats}
\usepackage{amsfonts}
\usepackage{mathrsfs}
\usepackage{color}
\usepackage[ruled]{algorithm2e}
\usepackage{framed}
\usepackage{psfrag, amssymb, amsmath, cite}
\usepackage{enumitem} 
\usepackage{capt-of}
\usepackage{hyperref}
\usepackage{url}
\usepackage{color}
\usepackage{gensymb}

\usepackage[pagewise]{lineno}
\DeclareMathOperator{\diag}{diag}

\newtheorem{lem}{\textbf{Lemma}}

\begin{document}

\title{Mapping Target Location from Doppler Data}

\author{Qingchen Liu,
		Samuel P. Drake and
        Brian D. O. Anderson

\thanks{
The work of Liu and Anderson was supported by the Australian Research Council (ARC) under grant DP-160104500 and by Data61-CSIRO.\newline
\indent Q. Liu and B.D.O. Anderson are with the Research School of Engineering, Australian National University, Canberra ACT 0200, Australia. {\tt\small \{qingchen.liu, brian.anderson\}@anu.edu.au}.\newline
\indent S. P. Drake is with the physics department at Adelaide University, also with the Defence Science and Technology (DST) group. {\tt\small samuel.drake@adelaide.edu.au}}}

       

\maketitle

\begin{abstract}
In this paper, we present an algorithm for determining a curve on the earth's terrain on which a stationary emitter must lie according to a single Doppler shift measured on an unmanned aerial vehicle (UAV) or a low earth orbit satellite (LEOS). The mobile vehicle measures the Doppler shift and uses it to build equations for a particular right circular cone according to the Doppler shift and the vehicle's velocity, then determines a curve consisting of points which represents the intersections of the cone with an ellipsoid that approximately describes the earth's surface. The intersection points of the cone with the ellipsoid are mapped into a digital terrain data set, namely Digital Terrain Elevation Data (DTED), to generate the intersection points on the earth's terrain. The work includes consideration of the possibility that the rotation of the earth could affect the Doppler shift, and of the errors resulting from the non-constant refractive index of the atmosphere and from lack of precise knowledge of the transmitter frequency. 
\end{abstract}

\section{Introduction}
 This paper is concerned with the following scenario. A stationary emitter is located at an unknown location on the surface of the earth. The frequency of its transmissions is known. An unmanned aerial vehicle (UAV) or a low earth orbit satellite (LEOS) receives the transmission. Using the Doppler shift, it is required to determine a curve, depicted using a 3D map tool, on which the transmitter must lie. 

An idealized version of this scenario might assume that the earth is a sphere, or possibly an ellipsoid, that all measurements (of position, velocity and frequency) were noise free and that the rotation of the earth can be neglected. In this case, the UAV or LEO receiver data serves to determine a right circular semi-infinite cone on which the emitter must lie, the axis coinciding with the velocity vector direction, the semi-angle derived from the Doppler shift, and the apex obtained from the receiver position. The intersection of this cone with the sphere/ellipsoid corresponding to the earth then provides the solution to the problem. 

There are however a number of complicating factors. These include 
\begin{enumerate}
\item
the multiple related but differing representations of the earth surface, including the WGS84 ellipsoid, defined by a constant gravitational equipotential surface; the EGM96 geoid, as an approximate ellipsoid with fine structure topographical variations capturing the height above sea level of points on the earth's surface \cite{wgs84}. 
\item
the different types of intersection/non-intersection that can occur between a cone and a ellipsoid. There may be no intersection; there may be a tangency; there may be a single curve; there may be two separate, i.e. non-intersecting,  curves (one of which may be in the 'shadow` of the earth as seen from the receiver), etc. 
\item
the data sources available to represent the earth,  e.g. the parameters of the WGS84 model \cite{wgs84}, the DTED data base \cite{DTED}, which indicates for samples spaced at a known interval the height of a point on the surface of the earth above the geoid. DTED is a matrix of terrain elevation values which provides basic quantitative data for systems and applications that require terrain elevation, slope, and/or surface roughness information. DTED data is uniformly spaced in angle (not distance).
\item
the fact that the earth is rotating about a polar axis, which means that given two points that are stationary with respect to the earth's surface except for the two poles, there is actually a relative velocity; roughly speaking this is a Coriolis effect. The value depends on the position of the two points. In consequence, it would seem that the determination of a cone using a Doppler shift, which amounts to indirect use of velocity data of the receiver (and assuming a stationary emitter), ought to take into account this relative velocity.
\item
the fact that a nominal rather than exact value for the emitter frequency may be known. (By way of example, a mobile phone tower may will an precisely known value, but a mobile phone may only have a nominal value). When a nominal value is used, there will be an error in the computation of Doppler shift and therefore of the cone and it is intersection with the Earth's surface is moved.   
\item
the fact that transmissions between a ground-located transmitter and a LEO receiver pass through a sufficiently long distance that an assumption of uniform value of the refractive index, or equivalently straight line propagation, is dangerous. (In contrast, if a UAV rather than LEO receives the transmission, the assumption of straight line propagation is almost certainly reasonable.) The refractive index varies with altitude. This variation will be significant between a ground receiver and a satellite, but minor between a ground receiver and a low-altitude UAV.
\end{enumerate}
A significant part of the contents of this paper seeks to deal, at least to some degree,  with these issues. 

Our work is of course not the first work on this subject. Prior contributions, some with more idealized assumptions than those we seek to have, certainly exist. In \cite{RN1152, RN1153, lee2007doppler}, geolocation of radio frequency (RF) emitters using Doppler shift measurements from mobile vehicles was considered. These works investigated scenarios assuming the vehicle path and the emitter are located in a plane, and so without considering the shape of the earth's surface. In \cite{ho1997geolocation}, a set of solutions from the combination of time difference of arrival (TDOA) and frequency difference of arrival (FDOA) measurements for localizing an emitter with known altitude above the earth surface was proposed. However, the method assumed an ellipsoid earth model without considering any height information. This restriction also appeared in \cite{pattison2000sensitivity, musicki2010mobile}, which had the similar problem settings to those in \cite{ho1997geolocation}.

 The structure of the paper is as follows. In the next section we provide a general review of various coordinate systems used for representing points on or above earth's surfaces, in this paper, relating them to the World Geodetic System (WGS) 1984 , and its related Geoid system for representing the earth's terrain. We introduce Digital Terrain Elevation Data (DTED), which is used as a data reference to index specific points on the earth's terrain. We also review standard coordinate systems that represent points on or above the earth surface, including earth-centered-earth-fixed coordinate systems, geographic coordinate systems and body coordinate systems, and explain their connections. In section III, we present fundamental Doppler-shift equations and identify two possible sources of Doppler shift in our scenario, viz. the motion relative to the earth's surface of the vehicle and the effect of the earth's rotation, and reveal the fact that the earth's rotation actually does not affect the Doppler shift, due to the fulfillment of a certain orthogonality condition. Though in our scenario, we consider a stationary emitter, the conclusion that the earth's rotation does not affect the Doppler shift is not dependent on this staitonarity, as the later derivation shows. We present the equations to build a particular right circular cone with the knowledge of the sensing vehicle position, velocity and the measured Doppler shift. In Section IV-A, we present an algorithm to find the intersections of a right circular cone with the WGS84 ellipsoid and identify different types of intersections. In Section IV-B, we present our developed method to find the cone-earth-terrain intersection curve where the emitter lies, by using the formerly built cone-earth-ellipsoid intersections and the DTED. We demonstrate the effectiveness of this method with some examples \footnote{Before moving to the next section, we refer the readers to the appendix for the notation and abbreviations used throughout this paper.}.
 
 
\section{Preliminaries}

\subsection{WGS84 ellipsoid and EGM96 geoid}
In general, global geodetic applications require three different surfaces to be clearly defined:
\begin{enumerate}
\item
The earth's topographic surface, which includes the landmass topography and the ocean bottom topography
\item
A geometrical or mathematical reference surface, which is an ellipsoid with known semi-axes.
\item
An equipotential surface, called the geoid. (Potential refers to gravity). Because of height variation, and because of density variation within the solid earth, this is bumpy rather than a smooth ellipsoid, and it does not coincide with the topographic surface. 
\end{enumerate}

The World Geodetic System (WGS) is a standard for use in cartography, geodesy, and satellite navigation including GPS. The latest revision is WGS84(G1762), established in 1984 and last revised in 2013. It comprises a standard coordinate system for the Earth, a standard ellipsoidal reference surface (the datum or reference ellipsoid) for raw altitude data, and a gravitational equipotential surface (the geoid) that defines the nominal sea level. 

The coordinate system defined in WGS84 is depicted in the next subsection.

The WGS84 reference ellipsoid is a mathematically defined surface that approximates the truer shape of the Earth. It is used as a preferred surface on which geodetic computations are performed and point coordinates such as latitude, longitude, and elevation are defined. The ellipsoid is defined in terms of the semi-major axis value $a=6378137.0\text{m}$ and a flattening coefficient $f=1/298.257223563$ (from which the semi-minor axis value $b=6356752.314245\text{m}$ can be determined). The fact that an ellipsoid rather than a sphere is involved gives rise to an important observation. Whereas in normal life, the notion of a vertical direction is considered to be identical with the notion of the direction of the force exerted by gravity on a body on or above the earth's surface (disregarding sign), these two directions cannot strictly be identified. 

The geoid is a smooth but highly irregular surface and has been built as a mathematical representation of the surface of the earth's gravity field. The geoid is widely used to describe mean sea level (MSL) since the surface of the gravity field coincides approximately with the mean sea level. Disregarding the terrain elevation on continents, the geoid is a much more accurate description of the true physical shape of the earth, than the WGS84 ellipsoid, though it is not of course identical. The WGS84 reference ellipsoid is the baseline of EGM96 geoid, i.e., the geoid undulations are with respect to the WGS84 ellipsoid. EGM96 geoid is a data grid overlay onto the WGS84 ellipsoid and can be used to find MSL for almost any point on the Earth at any given latitude and longitude.


\subsection {Coordinate systems}


In this subsection, we review standard material concerning the different coordinate systems used for representing points on or above the earth's surface 

The first of these systems is an \textit{earth-centered, earth-fixed (ECEF) cartesian coordinate system}. A stationary point on the surface of the earth has constant ECEF coordinates. The origin is taken to be the earth's center of mass. The $z$-axis passes from the origin through a point on the surface termed the reference pole (effectively the north pole), and the $x$-axis passes through a zero meridian (effectively the longitudinal line through Greenwich, UK). The $y$-axis is chosen to ensure a right-handed orthogonal coordinate system. 

The second coordinate system is the geographic coordinate system, which enables every location on Earth to be specified by a triple consisting of latitude $\phi$, longitude $\lambda$ and height $h$. The latitude of a point on or above the earth's surface (ellipsoid surface) is the angle between the equatorial plane and the normal at that point. Note that the normal at a point on an ellipsoid surface does not pass through the centre, except for points on the equator or at the poles. The height of a point is the vertical distance from the point to some surfaces (e.g. WGS84 ellipsoid or EGM96 geoid).

The third coordinate system is the body coordinate system, which is directly defined on the body of the vehicle. Its origin is located at the center of gravity of the vehicle. The $x$-axis of the body coordinate system points froward; the y-axis points to the right of the $x$-axis, perpendicular to the $x$-axis; the $z$-axis points down through the bottom the vehicle, perpendicular to the $x$-$y$ plane. 

The last coordinate system is the vehicle-carried navigation coordinate system, which is usually viewed as recording east, north and up displacements from some agreed point of origin. We denote the vehicle-carried navigation coordinate system with the term ENU coordinate system in this paper. Its origin is defined as the center of the gravity of the vehicle as well. A tangent plane is fitted to this fixed origin point, with the east, north and up axes pointing in the obvious directions.

\subsection{Connection between coordinate systems}
Navigation instruments such as the global positioning system (GPS) and the inertial navigation system (INS) may well determine location (latitude, longitude and height) and attitude (roll, pitch, yaw) in relation to some agreed vehicle and some system is needed to marry the use of measurements derived from such systems and ECEF measurements. 

We first consider the connection between ECEF and geographic coordinate systems. The output of a GPS receiver is latitude $\phi$, longitude $\lambda$ and height $h$ in the geographic coordinate system, denoted by $(\phi, \lambda, h)$ (the vertical baseline is the WGS84 ellipsoid) \cite{kaplan2005understanding}. The relevant equations for converting $(\phi, \lambda, \text{h})$ to $(x, y, z)$ in the ECEF coordinate system are  \cite{gerdan1999transforming}:
\begin{equation}
\begin{split}
x&=\left(\frac{a}{\chi}+h\right)\cos \phi\cos\lambda  \\
y&=\left(\frac{a}{\chi}+h\right)\cos \phi\sin\lambda  \\
z&=\left(\frac{a(1-e^2)}{\chi}+h\right)\sin\phi
\end{split}
\label{eq:geotocartesian}
\end{equation}
where $e^2=6.69437999014\times 10^{-3}$ is the square of the first eccentricity, and 
\begin{equation}
\chi=\sqrt{1-e^2\sin^2\phi}
\end{equation}\textbf{}
The inverse transformation from ECEF coordinates to geographic coordinates can be described by the following equations \cite{iliffe2000datums}:
\begin{equation}
\begin{split}
&\lambda = \arctan{\left(\frac{y}{x}\right)}\\
&\phi = \arctan\left(\frac{z(1-f)+(2f-f^2)a\sin^3\mu}{(1-f)(p-(2f-f^2)a\cos^3\mu)}\right)\\
&h = p\cos\phi + z\sin\phi - a\sqrt{1-(2f-f^2)sin^2\phi}
\label{eq:cartesiantogeo}
\end{split}
\end{equation}
where $p = \sqrt{x^2 + y^2}$, $r = \sqrt{p^2 + z^2}$, $f$ is the flattening of the ellipsoid and $\mu$ is a parameter calculated according to 
\begin{align*}
\mu = \arctan\left(\frac{z}{p}(1-f) + \frac{(2f-f^2)az}{rp} \right)\\
\end{align*}


We now present the connections between ECEF and ENU coordinate systems. Suppose that the origin for ENU coordinates is defined by a point $(x,y,z)$ in ECEF coordinates. Suppose that a point near to the specified point is defined by the coordinate differences $(dx,dy,dz)$ between the coordinates of the nearby point, viz, $(x+dx,y+dy,z+dz)$ and the coordinates of the specified origin point $(x,y,z)$. The ENU coordinates of the nearby point are $(de,dn,du)$, the origin point of course having ENU coordinates $(0,0,0)$. The question arises: how is $(de,dn,du)$ related to $(dx,dy,dz)$. This can be answered by understanding that the orientation of ENU coordinates is determined by rotating the ECEF coordinates; the first rotation is about the $z$-axis, by $\lambda$ degrees (corresponding to the longitude), and then rotating about the new $x$-axis (obtained from the old $x$-aixs through rotation of $\lambda$ degrees in the $x$-$y$ or equatorial plane) by $\phi$ degrees. Consequently we have
\resizebox{0.85\linewidth}{!}
{
\begin{minipage}{\linewidth}
\begin{eqnarray}
\left[\begin{array}{c}
de\\dn\\du
\end{array}\right]
&=&\left[\begin{array}{ccc}
1&0&0\\
0&-\sin\phi&\cos\phi\\
0&\cos\phi&\sin\phi\end{array}\right]
\left[\begin{array}{ccc}
-\sin\lambda&\cos\lambda&0\\
\cos\lambda&\sin\lambda&0\\
0&0&1
\end{array}\right]\left[\begin{array}{c}
dx\\dy\\dz
\end{array}\right] \nonumber \\
&=&\left[\begin{array}{ccc}
-\sin\lambda&\cos\lambda&0\\
-\sin\phi\cos\lambda&-\sin\phi\sin\lambda&\cos\phi\\
\cos\phi\cos\lambda&\cos\phi\sin\lambda&\sin\phi\end{array}\right]
\left[\begin{array}{c}
dx\\dy\\dz
\end{array}\right]
\label{eq:rotation_enu_ecef}
\end{eqnarray}
\end{minipage}
}

We now state the connections between the vehicle-carried navigation system and the body coordinate system. The output of INS is the Euler angles, the set of roll $\alpha$, pitch $\beta$, raw $\gamma$,  with respect to the vehicle-carried navigation coordinate system. The rotation matrix $\mathbf{R}^{\text{ENU}}_{\text{Body}}$ from the ENU coordinate system to the body coordinate system can be described by using the Euler angles, with the formula described by \eqref{eq:rotation_enu_body} at the top of the next page \cite{grewal2007global}: 

\newcounter{mytempeqncnt}
\begin{figure*}[t]
\normalsize
\setcounter{mytempeqncnt}{\value{equation}}
\setcounter{equation}{4}
\begin{align}
\mathbf{R}^{\text{ENU}}_{\text{Body}} = 
\begin{bmatrix}
\sin{\gamma}\cos{\beta}  &  \cos{\alpha}\cos{\gamma}+\sin{\alpha}\sin{\gamma}\sin{\beta}   &   -\sin{\alpha}\cos{\gamma} + \cos{\alpha}\sin{\gamma}\sin{\beta}  \\
\cos{\gamma}\cos{\beta}  &  -\cos{\alpha}\sin{\gamma}+\sin{\alpha}\cos{\gamma}\sin{\beta}   &   \sin{\alpha}\sin{\gamma} + \cos{\alpha}\cos{\gamma}\sin{\beta}  \\
\sin{\gamma}        &  -\sin{\alpha}\cos{\beta}                     &   -\cos{\alpha}\cos{\beta}                       \\
\end{bmatrix}\label{eq:rotation_enu_body}
\end{align}
\setcounter{equation}{\value{mytempeqncnt}}
\hrulefill
\vspace*{4pt}
\end{figure*}

\subsection{Dealing with height data}
Height data is available publicly and has been specified precisely in what is known as DTED (Digital Terrain Elevation Data) \cite{durland2009defining}. As noted earlier, DTED is a digital elevation model, consisting of a matrix of terrain elevation values, corresponding to the height of the ground above the geoid. The DTED format has three levels, termed level 0, level 1 and  level 2, corresponding respectively to  spacing of  approximately 900 meters, 90 meters and 30 meters. These line spacings correspond to specific arcsecond changes in latitude and longitude.

\begin{figure}
\centering
\includegraphics[width=80mm]{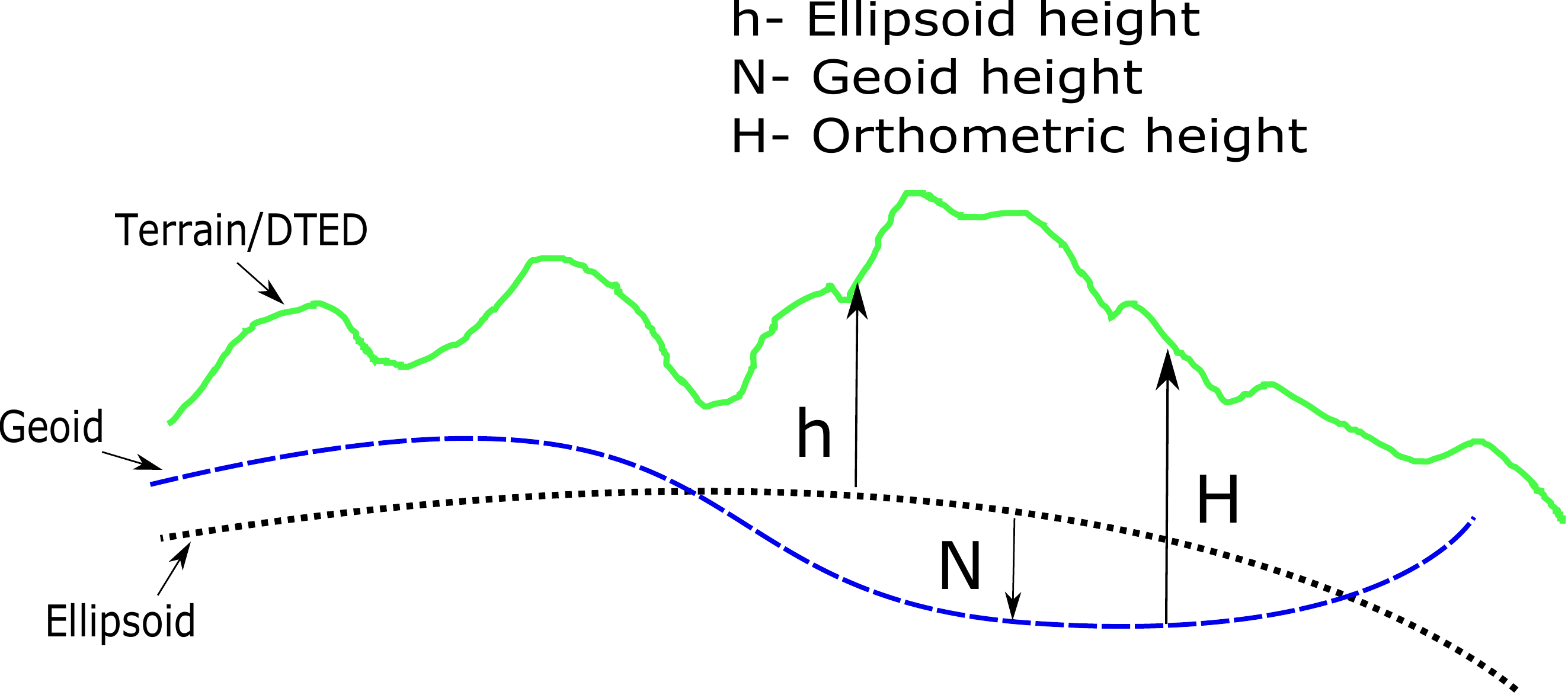}
\caption{\label{fig:geoid_height}The relationship between orthometric height, ellipsoid height and geoid height. Note that $N$ assumes a negative value when the geoid is below the ellipsoid. The figure is taken from \cite{durland2009defining}}
\end{figure}

DTED data is evidently indexed to specific points on the earth's terrain. The horizontal datum is referenced to the WGS84 ellipsoid and the vertical datum is referenced to the EGM96 geoid. For a particular point above the ground, there are three types of height: orthometric height is defined as the height of the terrain above the geoid: ellipsoid height is defined as the height of the terrain above the WGS84 ellipsoid; geoid height is defined as the height of the geoid above the  WGS84 ellipsoid. The relationship of orthometric height, ellipsoid height and geoid height is depicted in Fig. \ref{fig:geoid_height}, where an up arrow indicates a positive value and down arrow indicates a negative value. It is straightforward to see that the orthometric height at a given point with latitude $\phi$ and longitude $\lambda$ can be expressed by

\stepcounter{equation}
\begin{equation}\label{eq:transformation_h}
h(\phi,\lambda) = H(\phi,\lambda) + N(\phi,\lambda)
\end{equation}

\begin{figure}
\centering
\includegraphics[width=60mm]{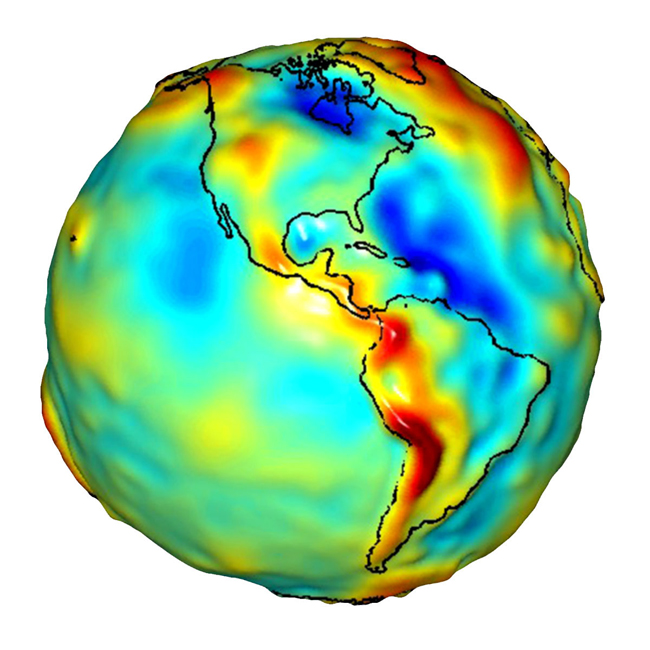}
\caption{\label{fig:geoid}A geoid. The figure is taken from \cite{durland2009defining}}
\end{figure}

\subsection{Problem statement}
Our task is one of passing from a set of measurements to a presentation in a useful (visual) form of information provided by those measurements. More specifically, our task is to define on a representation such as Google map a curve corresponding to those points consistent with a single FDOA measurement taken by an aircraft.

This task can be broken down as follows:

\begin{enumerate}
\item
Establish the equation of a right circular cone with fixed apex (corresponding to vehicle position), axis (corresponding to vehicle heading) and semi-angle (corresponding to FDOA measurement) in the ECEF coordinate system. 
\item
Establish a procedure for determining an array from which the curve of intersection of the two surfaces (the FDOA cone established in the first step and the WGS84 ellipsoid) can be constructed.
\item
Relate the ECEF coordinates to geographic coordinates.
\item
Make an adjustment to cope with ground height above sea level according to DTED.
\end{enumerate}

\section{Doppler shift and earth rotation}

\subsection{Locus of points of constant Doppler}

We now consider the problem of defining the possible emitter locations corresponding to a particular measured Doppler shift. The scenario is that there is a transmitter on the surface of the earth, and a receiver above the earth that can receive signals.

Throughout this section, we make several explicit assumptions. However, with warning, we will relax specific assumptions at certain points. The assumptions are:
\begin{enumerate}
\item
The earth's surface is modelled as the WGS84 ellipsoid
\item
A transmitter is located on the surface of the earth and a receiver mounted in a sensing vehicle is located above the surface of the earth, with the transmitter being visible from the receiver. 
\item
The transmitter is stationary, while the sensing vehicle is moving. 
\item
The velocity of light is constant and known; as a consequence, straight line propagation occurs. (This assumption is in a sense one of the least justifiable, and its relaxation will be explored later.)
\item
The exact position and velocity of the sensing vehicle are known at the time it measures a Doppler-shifted signal.
\item
The sensing vehicle can infer the Doppler shift associated with a received signal, because it has precise knowledge of the unshifted frequency at the transmitter.
%
%
%
\end{enumerate}

\begin{figure}[tb]
\centering
\includegraphics[width=80mm]{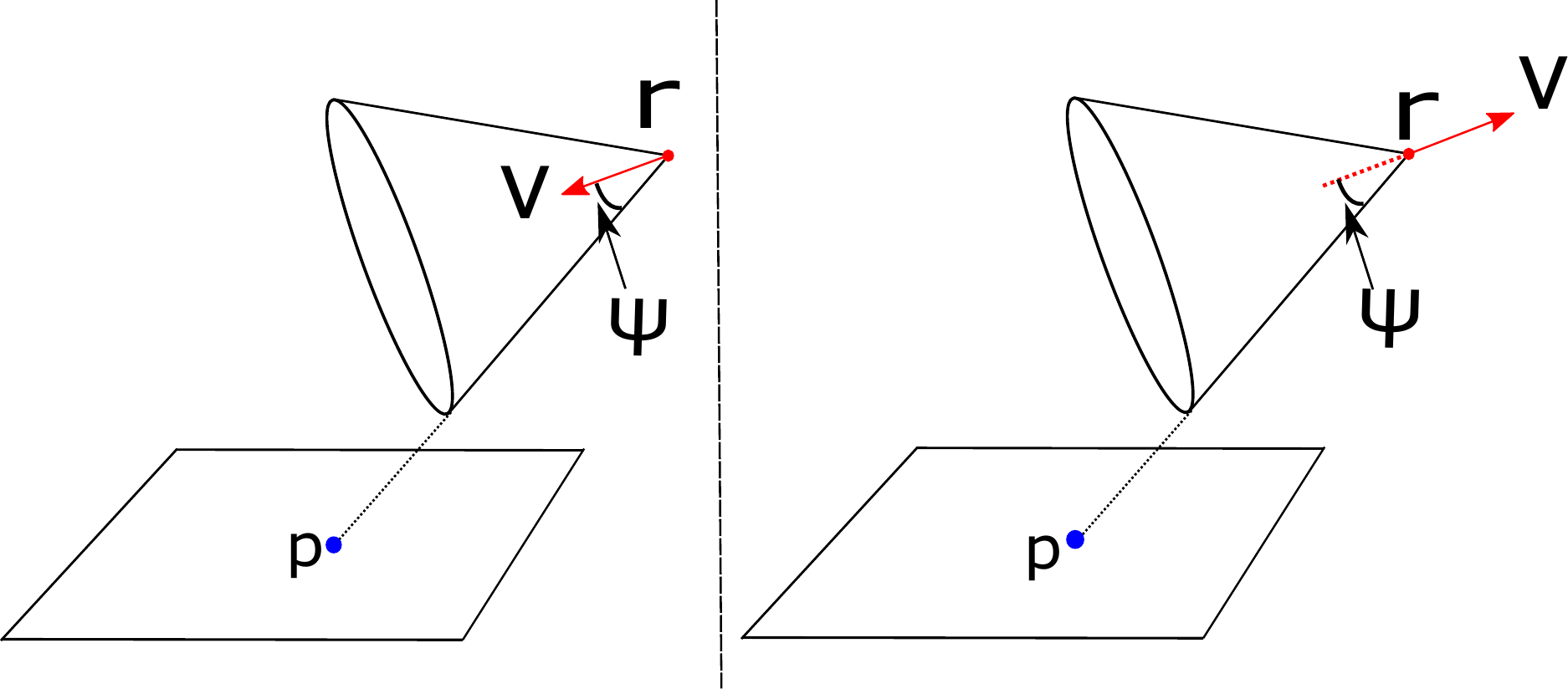}
\caption{\label{fig:fdoa_cone}The two cases of the FDOA Cone. The left-side figure indicates that the direction of axis of the FDOA cone has the same direction as the receiver velocity if the measured Doppler shift is positive. Otherwise, the direction of the axis of the FDOA cone is the reverse of the direction of velocity of the receiver, as illustrated in the right-side figure.}
\end{figure}

It is straightforward to make a general statement about the nature of the locus of the points corresponding to a particular Doppler shift. Knowledge of the Doppler shift associated with a received signal implies knowledge of a circular cone on which the transmitter must lie. The cone's apex is at the sensing vehicle position, and axis aligned with the velocity vector of the sensing vehicle. The direction of the axis can be either the same or reverse of the velocity direction, depending on the sign of the measured Doppler shift. If the measured Doppler shift is positive, then the axis of the cone is defined with the same direction of the velocity vector, otherwise the direction of axis is the reverse of the velocity vector, see Fig. \ref{fig:fdoa_cone}. The semi-angle, call it $\psi$, can be computed from knowledge of the magnitude of the velocity $|\mathbf{v}|$ of the sensing vehicle, the speed of light $c$, the unshifted transmitter frequency $f_0$, and the Doppler shift $\delta$: \footnote{The transmitter is assumed to be stationary, and for the moment we neglect any contribution to the relative velocity of the receiver with respect to the transmitter due to the earth's rotation. This will be dealt with subsequently. }

\begin{equation} \label{eq:cone_semi_angle}
\cos \psi=(\delta/f_0)/(|\mathbf{v}|/c)
\end{equation}
The fractional Doppler shift equals the fraction relative to the speed of light of the component of the sensing vehicle's velocity along the direction to the transmitter. 

The FDOA cone and WGS84 ellipsoid are both two-dimensional surfaces in a 3D space. They intersect in at least one point, namely the point where the transmitter is located. Generically, if there is a nonempty intersection, it will be a one-dimensional smooth set, a curve in space in fact. Exceptionally, as a kind of limiting case, the intersection may be a single point. This would arise if the tangent plane to the ellipsoid coincided with the tangent plane to the cone at the common point of intersection.

Our ultimate goal is to the intersection curve. 

\subsection{Coriolis Effect}
In this subsection, we highlight the need to consider the fact that the earth is rotating, when considering the determination of the cone associated with a Doppler shift. We demonstrate that the operative kinematics constraints imply that the relative velocity component between an emitter and a receiver due to the earth's rotation makes zero contribution to the Doppler shift.


Viewed from a point in a coordinate frame that is not earth fixed, but which views the earth as rotating, any two points on the earth's surface (other than the poles) can be seen to have velocities associated with the rotation of the earth about its axis. Further, any two distinct points excluding the two poles have \textit{different} velocities on this account. (The yearly motion of the earth around the sun is discounted in making these observations.) It follows that any two points on (or above) the earth's surface which in an ECEF coordinate frame is concerned would be considered stationary actually have a relative velocity (whose magnitude can be substantial, e.g. some tens of m/sec), and thus potentially gives rise to a Doppler shift between the transmit/receive frequency of an RF signal propagating from one point to the other. Further, if one of the points is fixed on or above the earth's surface, and the other is moving (relative to the earth's surface), the relative velocity can be regarded as coming from two causes, the motion relative to the earth's surface of one point, and the effect of the earth's rotation (`Coriolis effect').  It is because a coordinate frame fixed to the earth is actually rotating, albeit with a fixed angular velocity,  rather than just translating at a uniform velocity, that this Coriolis effect arises. 

We now establish how the relative velocity can be determined, and seek to determine the associated Doppler shift. For this purpose, it is the case to attribute the Coriolis component to rotation of an earth-centred earth-fixed coordinate basis.

In ECEF coordinates we denote the emitter coordinates by using a vector $\mathbf{p}:=p^x \mathbf{e}_x+p^y\mathbf{e}_y+p^z\mathbf{e}_z$ and the receiver coordinates by $\mathbf{r}:=r^x\mathbf{e}_x+r^y\mathbf{e}_y+r^z\mathbf{e}_z$. Where $\mathbf{e}_x,\mathbf{e}_y,\mathbf{e}_z$ denote unit vectors aligned with the ECEF coordinate axes. Due to the rotation of the earth, these vectors are not stationary, when seen from an inertial frame. The transmitter frequency $f_r$ measured at the receiver is 
\begin{equation}\label{eq:generalDoppler}
f_r = f_0 \left(1- \frac{1}{c}\left[\frac{d(\mathbf{p}-\mathbf{r})}{dt}\right]\cdot\frac{\mathbf{p}-\mathbf{r}}{\|\mathbf{p}-\mathbf{r}\|}\right)
\end{equation}
In computing the derivative, we must allow for motion relative to the ECEF system (values of the coordinates of $\mathbf{p},\mathbf{r}$ change) and, separately, motion of the coordinate axes. 

In the ECEF coordinate system the basis vectors rotate so that \cite{goldstein2002classical}
\[
\frac{d\mathbf{e}_i}{dt} = \mathbf{\Omega}\times \mathbf{e}_i\;\;i\in\{x,y,z\} \quad, 
\]
 where $\Omega$ is the magnitude of the angular velocity. Consequently the relative velocity between the emitter and the receiver is the sum of the coordinate velocity and the Coriolis effect: 
\begin{eqnarray}
\frac{d\left(\mathbf{p} - \mathbf{r}\right)}{dt} &=& 
\sum_{i\in\{x,y,z\}}\frac{d\left(p^i-r^i\right)}{dt}\mathbf{e}_i + \left(p^i-r^i\right)\mathbf{\Omega}\times \mathbf{e}_i  \nonumber \\
&=&  \left( \sum_{i} \frac{d \left(p^i-r^i\right)}{dt} \mathbf{e}_i\right) + \mathbf{\Omega}\times\left(\mathbf{p}-\mathbf{r}\right)
\label{eq:dpr}
\end{eqnarray}
The second term on the right hand side of~\eqref{eq:dpr} is due to the Coriolis effect. Note that this vector is perpendicular to $\mathbf{p}-\mathbf{r}$ and hence doesn't contribute to the Doppler effect:
\begin{align*}
    f = f_0 \left( 1 - \frac{1}{c} \left( \sum_i \frac{d \left(p^i -r^i\right)}{dt} \mathbf{e}_i \cdot \frac{\mathbf{p}-\mathbf{r}}{\|\mathbf{p}-\mathbf{r} \|}\right)\right)
\end{align*}
This is exactly the formula we would write down if we failed to take into account the rotation of the earth, or put another, we can simply neglect the effect associated with rotation of the earth.

\subsection{Equation of an arbitrary right circular cone}
In this section, our aim is to find an equation representing a right circular cone with apex coordinates $(r_x,r_y,r_z)$ and unit vector in the direction of velocity given by $(\alpha,\beta,\gamma)$, i.e. subject to $\alpha^2+\beta^2+\gamma^2=1$, in ECEF coordinate system. Note that a mobile vehicle can determine its attitude $(\alpha,\beta,\gamma)$ with the help of the INS devices. The velocity direction vector $(\alpha,\beta,\gamma)$ cannot be directly obtained but can be calculated by transforming a unit vector $(1,0,0)$ from the vehicle's body coordinate system to the ECEF coordinate system by using the inverse of the rotation matrices provided in \eqref{eq:rotation_enu_ecef} and \eqref{eq:rotation_enu_body}.

Our starting point is that, as is well-known and indeed easily checked,  (the surface of) a cone with axis corresponding to the $z$-axis and apex at the origin is given by an equation of the form

\begin{equation}\label{eq:original_cone}
\frac{x^2}{d^2}+\frac{y^2}{d^2}-z^2=0
\end{equation}

The semi-angle is $\tan^{-1}d$.

Our immediate goal now is to understand how to handle translation of the apex and rotation of the axis of the cone. If the apex of the cone is at $(r_x,r_y,r_z)$, then the above equation is replaced by
\begin{equation}
\frac{(x-r_x)^2}{d^2}+\frac{(y-r_y)^2}{d^2}-(z-r_z)^2=0
\end{equation}
To understand how to handle rotation, we first observe a simple Lemma. 
\begin{lem}
Let $\alpha,\beta,\gamma$ be a set of direction cosines of a real 3-vector, i.e. $\alpha^2+\beta^2+\gamma^2=1$. Then the following matrix is an orthogonal rotation matrix:
\begin{equation}\label{eq:rotation_matrix}
\mathbf{R}=\left[\begin{array}{ccc}
\frac{\alpha\gamma}{(\alpha^2+\beta^2)^{1/2}}&-\frac{\beta}{(\alpha^2+\beta^2)^{1/2}}&\alpha\\
\frac{\beta\gamma}{(\alpha^2+\beta^2)^{1/2}}&\frac{\alpha}{(\alpha^2+\beta^2)^{1/2}}&\beta\\
-(\alpha^2+\beta^2)^{1/2}&0&\gamma\end{array}
\right]
\end{equation}
\end{lem}
By way of an outline proof, we observe first that it is easily verified that each of the columns is a vector of length 1, and the columns are mutually orthogonal. Further, a straightforward calculation shows that the determinant is 1, assuring that the matrix is a rotation matrix.

When $\alpha=\beta=0$, four entries of the matrix are not well-defined, and so the continuity of the matrix comes into question. However, let $\theta$ be such that $\alpha=\sqrt{1-\gamma^2}\cos\theta$ and $\beta=\sqrt{1-\gamma^2}\sin\theta$. Note that given the direction cosines $\alpha, \beta$ and $ \gamma$, such a $\theta$ always exists and is unique. Let us regard the matrix $R$ as a function of $\theta$ and $\gamma$. Then we can write
\begin{equation}
\mathbf{R}(\theta,\gamma)=\left[\begin{array}{ccc}
\gamma\cos\theta&-\sin\theta&\sqrt{1-\gamma^2}\cos\theta\\
\gamma\sin\theta&\cos\theta&\sqrt{1-\gamma^2}\sin\theta\\
-\sqrt{1-\gamma^2}&0&\gamma\end{array}\right]
\end{equation}
Now if $\alpha,\beta$ tend continuously to zero  while obeying $\alpha^2+\beta^2<0$ (except in the limit), i.e. $\gamma<1$ except in the limit, and if they tend to zero in such a way that $\beta/\alpha$ also approaches a limit, it is evident that $R(\theta,\gamma)$ will approach the limit
\begin{equation}
\mathbf{R}(\theta,1)=\left[\begin{array}{ccc}
\cos\theta&-\sin\theta&0\\\sin\theta&\cos\theta&0\\
0&0&1
\end{array}
\right]
\end{equation}

Now suppose a right-circular cone with apex at the origin is such that the axis of the cone has direction cosines $\alpha,\beta,\gamma$. Assume temporarily that a coordinate basis with coordinates $\bar{\mathbf{p}}=[\bar x ~ \bar y ~ \bar z]^\top$ is established with the same origin and with the $\bar z$-axis coinciding with the axis of the cone. The equation of the cone in the new coordinate basis is given by
\begin{equation}\label{eq:standard cone}
\frac{\bar x^2}{d^2}+\frac{\bar y^2}{d^2}-\bar z^2=0
\end{equation}
for suitably chosen $d$. Write this equation as 
\begin{equation}\label{eq:niceaxiscone}
\bar{\mathbf p}^{\top}\bar \Lambda\bar{\mathbf p}=0
\end{equation}
where 
\begin{equation}
\bar\Lambda=\mbox{diag}[d^{-2},d^{-2},-1]
\end{equation}
It is clear that we want correspondence between the line $\bar x=0, \bar y=0$, i.e. the axis of the cone in $\bar x, \bar y, \bar z$ space and the axis of the cone in $x,y,z$ space, which is defined by direction cosines $\alpha,\beta, \gamma$. Because of the structure of the last column of $\mathbf{R}$, this correspondence is assured if we take $\mathbf{p}=\mathbf{R}\bar{\mathbf{p}}$ or
\begin{equation}
\left[\begin{array}{c}
x\\y\\z\end{array}\right]=\mathbf{R}\left[\begin{array}{c}
\bar x\\ \bar y\\ \bar z
\end{array}\right]
\end{equation}
Equivalently, the equation in $x,y,z$ space of the cone follows from \eqref{eq:niceaxiscone} as 
\begin{equation}
\mathbf{p}^{\top}\mathbf{R}\bar\Lambda \mathbf{R}^{\top}\mathbf{p}=0
\end{equation}

The matrix $\mathbf{R}(\theta,1)$ is relevant in a special case. If $\alpha=0,\beta=0, \gamma=1$ define the directions in which the axis of the cone should point, then it is evident that the original cone, with $z$-axis coinciding with the direction of the cone is already correctly positioned. Thus one might reasonably suppose that the transforming rotation matrix should be the identity matrix. The formula for $\mathbf{R}(\theta,1)$ defines a rotation matrix which leaves that axis invariant, as it causes rotation only of the $x,y$ coordinate plane about the $z$-axis, which is normal to it.

Combining the results on translation and rotation, we have the following Lemma. 
\begin{lem}
Consider a right circular cone with apex at $\mathbf{p}=[r_x,r_y,r_z]^{\top}$, with axis given by the unit norm vector of direction cosines $[\alpha,\beta,\gamma]^{\top}$ and with semi-angle given by $\tan^{-1}d$. Then the equation of this cone is 
\begin{equation}
\left[x-r_x\;y-r_y\;z-r_z\right]\mathbf{R} \diag[d^{-2},d^{-2},-1]\mathbf{R}^{\top}\left[\begin{array}{c}
x-r_x\\
y-r_y\\
z-r_z
\end{array}\right]=0
\label{eq:transformed_cone}
\end{equation}

\end{lem}

Our next major task is to study the intersection of this conical surface with the ellipsoid defined by the earth.

\section{Determining the intersection of the FDOA cone with the earth's terrain}

Here is a summary of what is required. 

The equation of WGS84 ellipsoid (with the positive $z$-axis identified with a line starting at the center of the earth and passing through the north pole) in ECEF coordinates is
\begin{equation}\label{eq:earth_ellipsoid}
\frac{x^2}{a^2}+\frac{y^2}{a^2}+\frac{z^2}{b^2}=1
\end{equation}
where $a$ represents semi-major axis value and $b$ represents semi-minor axis value.

One seeks to determine the intersection between an arbitrary right circular cone (with parameters corresponding to typical values obtained from a UAV or LEOS) and the WGS84 ellipsoid.

This intersection will in general be a curve, or more accurately, may comprise one or more disjoint curves. To see that for example two (non-intersecting) curves could be obtained, suppose that the apex of the cone lay outside the ellipsoid, say above the north pole, with the axis of the cone passing through the centre of the ellipsoid. Then if  the semiangle of the cone were sufficiently small, the cone would  intersect the upper hemisphere of the ellipsoid in a closed curve, and separately it would intersect the lower hemisphere of the ellipsoid in a second closed curve. Of course, this is not the only possibility.

For the purposes of establishing and recording the intersection of the cone and the ellipsoid, one could imagine choosing a set of neighboring values of $x$ (or $y$ or $z$), and for each value determining the real intersection points (which will normally be 0, 2 or 4 in number) of the two equations. (We say normally, because the number may be different if there is a point at which the two surfaces touch and have some sort of common tangency.) 

Given the possibility of the intersection comprising two separate curves, it would be important to determine, for neighboring values of $x$, which intersection points were on the same curve, and which were not. When the intersection points are remote from one another, this will be easy. It will be less straightforward if curves merge, touch, etc. 


\subsection{Find the intersections of a cone with an ellipsoid}
To parametrize a right circular cone, we can divide the cone surface into two parts, an apex $\mathbf{r}=(r_x, r_y, r_z)$ and a set $\Phi$ of straight rays which start from the apex and ends in the infinity. Suppose $\alpha_c$, $\beta_c$ and $\gamma_c$ are direction cosines of one of the rays in $\Phi$. The ray equation can be given by
\begin{equation}\label{eq:cone_para}
\begin{split}
x = r_x + \alpha_c s\\
y = r_y + \beta_c s\\
z = r_z + \gamma_c s\\
\end{split}
\end{equation}
where $s\in[0, \infty)$ is the range of the ray. The constraint of non-negative value on $s$ ensures that \eqref{eq:cone_para} describes a right circular cone rather than a double cone.

\begin{figure}
\centering
\includegraphics[width=60mm]{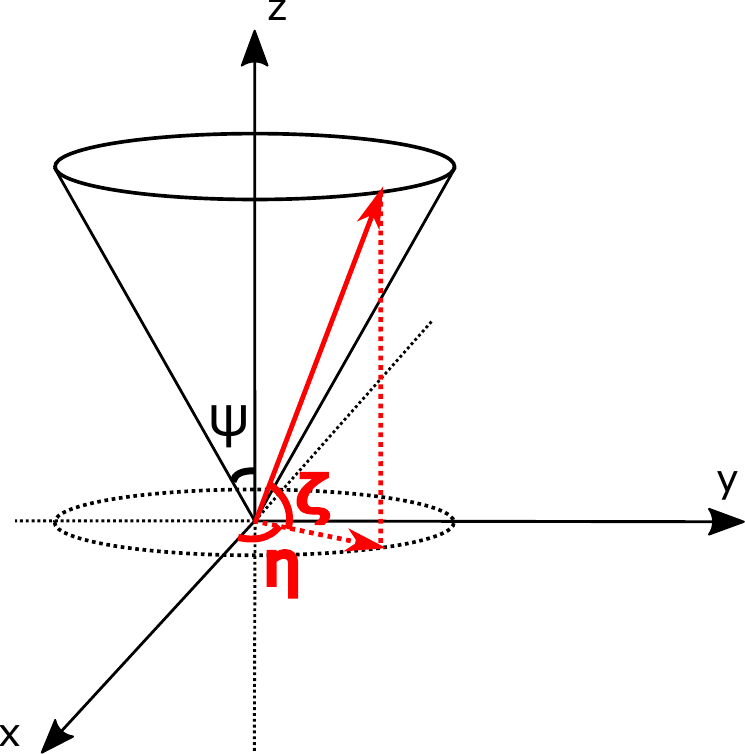}
\caption{\label{fig:cone_coordinates}Illustration of $\zeta$ and $\eta$.}
\end{figure}

Now the problem of finding the intersections of two quadratic surfaces (a right circular cone and an ellipsoid) has been transformed to finding the intersections of a set of rays described by \eqref{eq:cone_para} with the ellipsoid \eqref{eq:earth_ellipsoid}. According to \eqref{eq:cone_para}, it is observed that the direction cosines $\alpha_c$, $\beta_c$ and $\gamma_c$ can be obtained by transforming the direction cosines of the rays on the surface of the original cone \eqref{eq:original_cone}. We first pick one ray on the surface of the cone \eqref{eq:original_cone} and identify its direction cosines $\alpha_r$, $\beta_r$ and $\gamma_r$ as follows:
\begin{equation}
\begin{split}
\alpha_r &= \cos(\zeta)\cos(\eta)\\
\beta_r  &= \cos(\zeta)\sin(\eta)\\
\gamma_r &= \sin(\zeta)\\
\end{split}
\end{equation}
where $\zeta$ denotes the angle from the ray to its projection on $x-y$ plane, and $\eta$ denotes the angle between the $x-{\text{axis}}$ and the projection of the ray on $x-y$ plane. See Fig. \ref{fig:cone_coordinates} for the illustrations of $\zeta$ and $\eta$. Let $\mathbf{d}_r: = [\alpha_r ~  \beta_r ~ \gamma_r]^\top$ be an unit direction vector of an arbitrary ray on cone \eqref{eq:original_cone}. The unit direction vector $\mathbf{d}_r: = [\alpha_t ~  \beta_t ~ \gamma_t]^\top$ of the mapping of the ray on the transformed cone $\eqref{eq:transformed_cone}$ can be directly calculated by 
\begin{equation}
\mathbf{d}_t = \mathbf{R}\mathbf{d}_r.
\end{equation}
Here $\mathbf{R}$ is the rotation matrix defined in \eqref{eq:rotation_matrix}.

Then for any particular ray, $r_x, r_y, r_z, \alpha_c, \beta_c$ and $\gamma_c$ are known. By substituting \eqref{eq:cone_para} into the ellipsoid equation \eqref{eq:earth_ellipsoid}, we have a second-order equation in $s$ with the form
\begin{equation}\label{eq:s}
a_s s^2 + b_s s + c_s = 0
\end{equation}
where
\begin{equation}\label{eq:abc}
\begin{split}
a_s &= q_1\alpha_c^2 + q_2\beta_c^2 + q_3\gamma_c^2\\
b_s &= 2q_1r_x\alpha_c + 2q_2r_y\beta_c + 2q_3r_z\gamma_c\\
c_s &= q_1r_x^2 + q_2r_y^2 + q_3r_z^2 + q_0;\\
\end{split}
\end{equation}
are all known parameters with $q_0 = -1$, $q_1=q_2=1/a^2$ and $q_3 = 1/b^2$. The real, non-negative solutions of \eqref{eq:s} define points on the ray corresponding to its intersections with ellipsoid. Note that the number of solutions of \eqref{eq:s} can be zero, one or two, depends on the value of the "quadratic discriminant":
\begin{equation}
D_s = b_s^2-4a_s c_s
\end{equation}
The number of solutions of \eqref{eq:s} also represents the number the intersection points of the ray with the ellipsoid.

\subsection{Find the intersections of the FDOA cone with the earth's terrain}
The aim of this subsection is to extend the calculation of the previous subsection involving the WGS84 ellipsoid to provide an algorithm to find the intersections of the FDOA cone with the earth terrain and plot these intersection points on a 3D map tool (e.g. google earth). We assume that the vehicle's measurements of its position, velocity and the Doppler shift are all accurate, i.e. without error.

Since the earth terrain is uneven, we cannot find equations to describe the terrain, which means we cannot directly calculate the intersections of a cone with the terrain by solving mathematical equations. Motivated by the observations that the earth terrain can be modelled by the discrete point set DTED, our idea is to find the intersection of the cone with the WGS84 ellipsoid and map these intersection points to the DTED dataset, according to the rays generated between the cone apex and the intersection points on the ellipsoid surface.  

Our algorithm is consisted of the following steps: 
\begin{enumerate}
\item Introduce $\mathrm{D}$ as a set of points comprising the DTED. Thus
\begin{align*}
\mathrm{D} := \{(\phi_i,\lambda_i,H_i): (\phi_i,\lambda_i,H_i) ~\text{is in the DTED},\notag\\
i=1,2,\ldots\}     
\end{align*}
Note that $\phi_i$ and $\lambda_i$ are defined with respect to the WGS84 ellipsoid, while $H_i$ is defined with respect to the EGM96 geoid \cite{durland2009defining} . For further use, we need to transfer $H_i$ (orthometric height) to $h_i$ (ellipsoid height) by using \eqref{eq:transformation_h}. We denote this new set as $\hat{\mathrm{D}}$.

\item By applying $\eqref{eq:geotocartesian}$ on each point in $\hat{\mathrm{D}}$, we obtain a new set $\hat{D}_c$, in which the coordinates for each point are defined with respect to the ECEF Cartesian coordinate system.

\item By using the method introduced in the previous subsection, the vehicle determines a set of the intersection points of the FDOA cone (associated with one single measurement) with the WGS84 ellipsoid. Introduce 
\begin{equation*}
\mathrm{A} := \{(x_i,y_i,z_i): (x_i,y_i,z_i)~\text{is on the intersection curve.}\} \end{equation*}
to denote the set of intersection points. Note that the coordinates associated with the points in $\mathrm{A}$ are defined with respect to the ECEF Cartesian coordinate system.

\item Select a point $p_{\mathrm{A}}^i$ in $\mathrm{A}$. By associating the coordinates of $p_{\mathrm{A}}^i$ with the receiver coordinates $\mathbf{r}$, we can define a line described by the following equation
\begin{equation}
\frac{x_i - r_x}{a_i} = \frac{y_i - r_y}{b_i} = \frac{z_i - r_z}{c_i}    
\end{equation}
where $a_i, b_i, c_i$ can be obtained by substituting the coordinates of $p_A^i$ and $\mathrm{r}$ into the above equation. For later use, we denote this line as $l_i$. It is one of the rays that comprise the surface of the FDOA cone.

\item Now we have a transformed DTED point set $\hat{\mathrm{D}}_c$ and the line $l_i$. For each point $p_k, k=1,2,\ldots$ in $\hat{\mathrm{D}}_c$, we can calculate its Euclidean distance $d_k^i$ to $l_i$.  We then select a fixed threshold $tr_i$ (the threshold is determined according to the spacing accuracy of the DTED sets) and figure out a new point set $\hat{D}_i$, defined as 
\begin{equation}
\hat{D}_i := \{(x_k,y_k,z_k): (x_k,y_k,z_k)\in\hat{D}_c, d_k^i\leq tr_i\}     
\end{equation}
The mapping of $p_A^i$ onto $\hat{D}_c$ is thus included in $\hat{D}_i$.

\item For each point $\hat{D}_i$, we calculate its Euclidean distance to the receiver, the point $\tilde{p}$ associated with the minimum distance to the receiver is the mapping point of $p_A^i$ onto $\hat{\mathrm{D}}_c$.

\item By repeating the steps 4)-6) for each $l_i$, we can finally find the cone-terrain intersection points.

\item For each cone-terrain intersection point, we use \eqref{eq:cartesiantogeo} to transform its ECEF coordinates into the geographical coordinates to depict it in a 3d map tool. 

\end{enumerate}

\begin{figure*}[t]
\centering
\includegraphics[width=170mm]{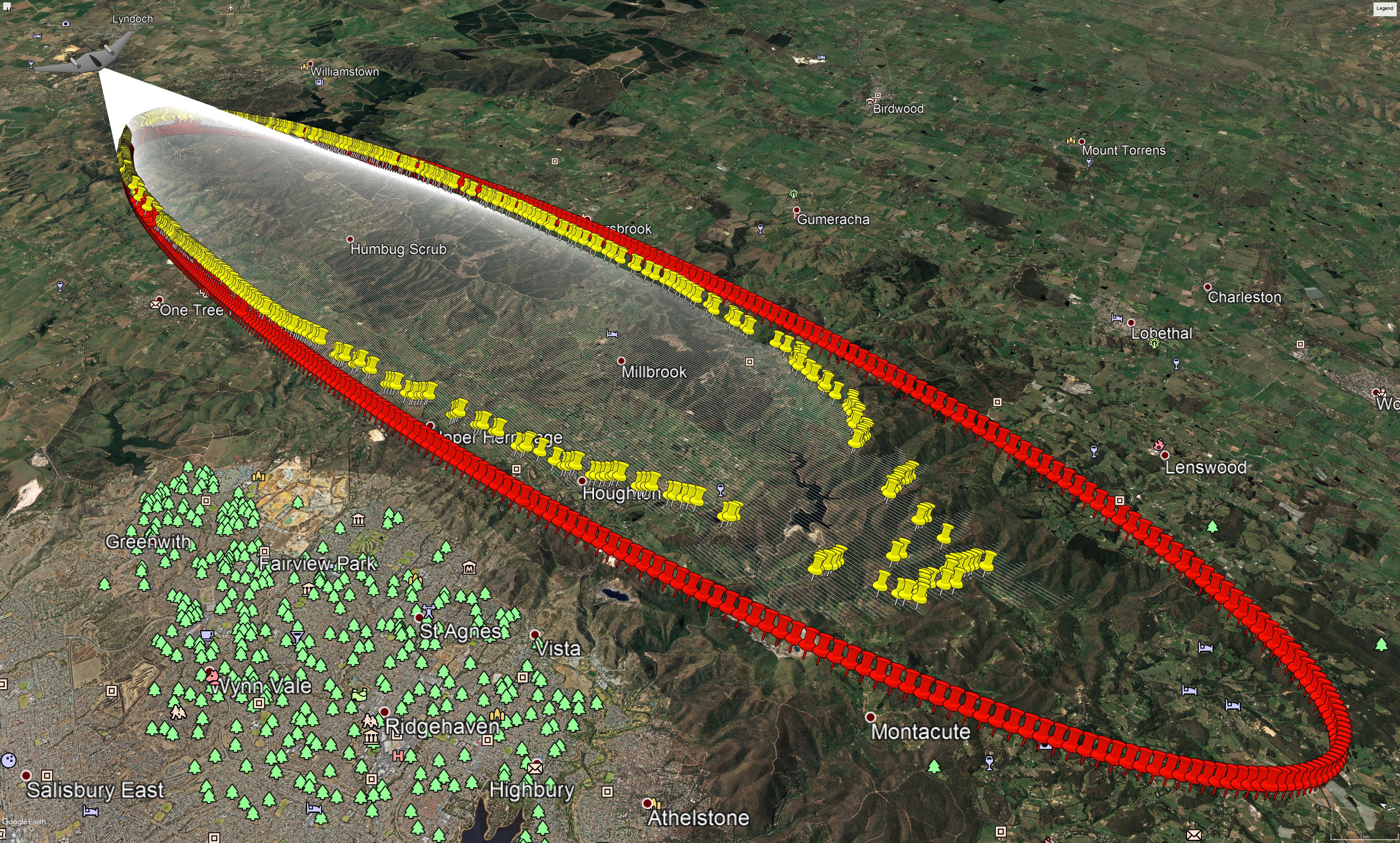}
\caption{\label{fig:simulation_1}The performance of the method for finding the intersections of the FDOA cone with the earth's terrain. In the simulation, the instantaneous UAV parameters are selected as: roll \ang{0}, pitch \ang{-30}, yaw \ang{190}, latitude \ang{-34.6462}, longitude \ang{138.833} and height 2000m. The semi-angle of the FDOA cone is assumed to be \ang{26.56}. The white rays represent the surface of the FDOA cone. The yellow place-marks indicate the intersections of the FDOA cone with the earth's terrain. The red place-marks represent the intersections of the FDOA cone with the WGS84 ellipsoid. Note that these red place-marks should be under the earth terrain since their ellipsoid heights are zero. The ellipsoid heights of the cone-ellipsoid intersection points are increased manually so that we can compare the difference between the cone-terrain intersections and the cone-ellipsoid intersections.}
\end{figure*}

\begin{figure*}[t]
\centering
\includegraphics[width=170mm]{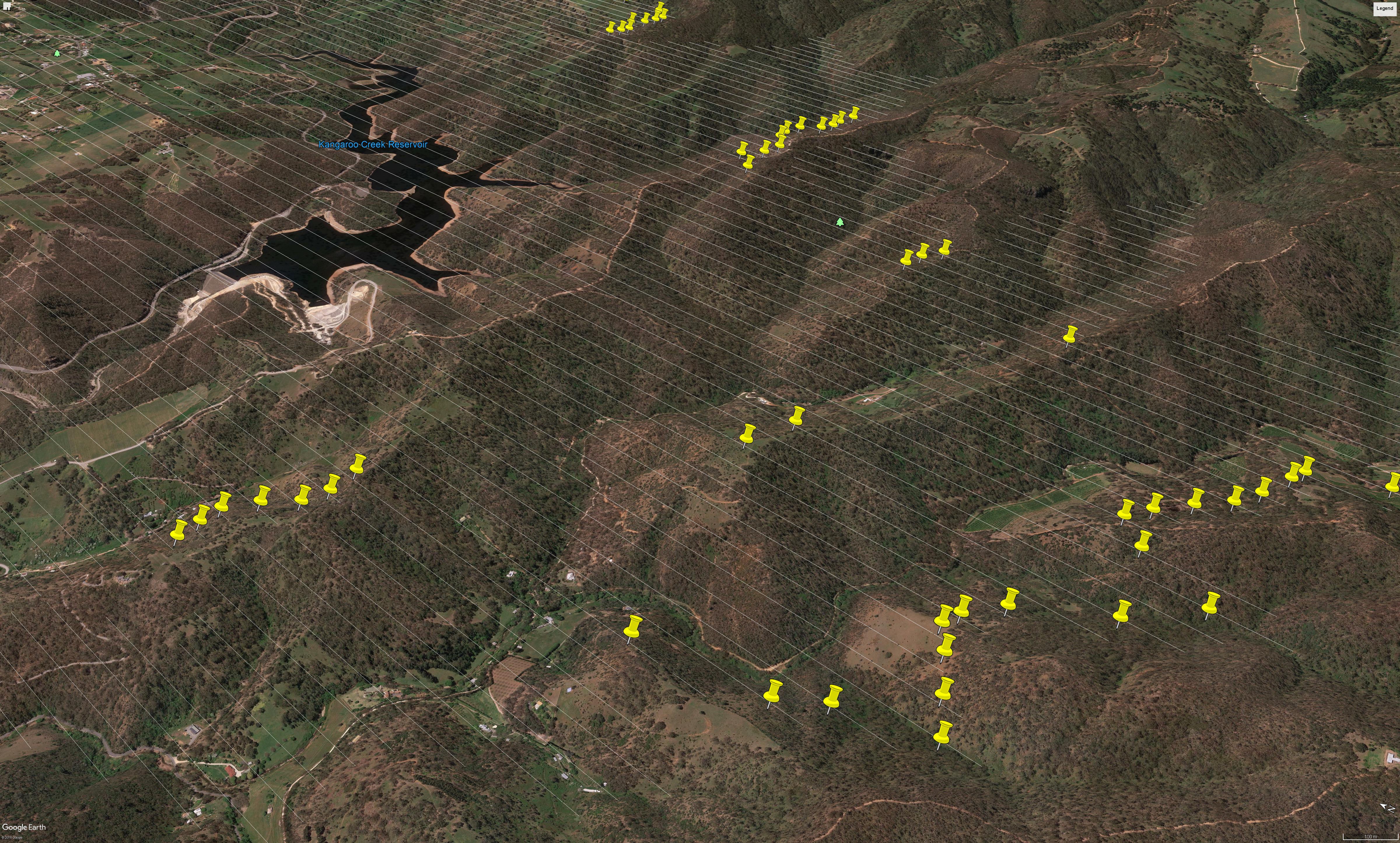}
\caption{\label{fig:simulation_1_intersection}\color{blue}The intersection points on the terrain.}
\end{figure*}

In the following, we use an example associated with UAV cases to illustrate the performance of our algorithms. The UAV position is selected to be close to the city of Adelaide, Australia. The FDOA cone, the intersections of the FDOA cone with the earth terrain, and the UAV are depicted in Fig. \ref{fig:simulation_1}. We also provide Fig. \ref{fig:simulation_1_intersection} for the purpose of clear observations on how the intersection points are distributed on the earth's terrain.


\section{Error analysis}
This section aims to discuss the curve shift associated with different errors raised in practical scenarios. We identify two potential error sources, 1) the fact that the nominal frequency of the emitter may be different from its true frequency; 2) the changing refractive index of the atmosphere, which results in a variation of the speed of the electromagnetic wave. We provide numerical examples to illustrate the curve shifts associated with different errors. Since UAV and LEOS are operated in different environments, we divide the discussions into two cases. For LEOS case, we further consider the relativistic Doppler effect. For all the examples in this section, we assume that the measured frequency at the receiver and the navigation (position, velocity and attitude) data of the vehicle are accurate, i.e. without errors. Note that for calculation simplicity, we assume that the nominal frequency of the emitter is 299,792,458$\text{Hz}$.

\subsection{Error sources}
We now discuss the first error source. The carrier frequency of the emitter is generated according to its built-in oscillators. The output frequency of an oscillator has a drift due to its internal changes with respect to time plus changes in the environment. So there is potentially an offset between the nominal frequency and the true frequency of the emitter. Before this section, we assumed that the true frequency rather than the nominal frequency of the emitter is precisely known to the receiver. This can only be assumed if, for example, there is a separate stationary receiver providing the information to the moving receiver, or the emitter is actually reflecting signals due to its being illuminated by a radar, with the position, velocity and precise frequency of the radar transmitter known to the moving receiver.

We then explain the second error source. It is well-known that the speed of electromagnetic wave in a medium different to vacuum, is different to its speed in the vacuum. Since the speed of electromagnetic wave changes with respect to the air refractive index when travelling in the air, a small error will be introduced when calculating the semi-angle of the FDOA cone using \eqref{eq:cone_semi_angle}. The speed of electromagnetic wave in the air is normally calculated using:
\begin{equation}
c_a = c/n
\label{eq:light_speed_in_air}
\end{equation}
where $n$ is the refractive index of the air. The refractive index of the air varies according to air temperature and pressure, which leads to a refractive index gradient in the atmosphere \cite{edlen1966refractive}. Note that air temperature and pressure can also be described with respect to the altitude. In this paper, we refer to an air refractive index model described by a function of altitude proposed in \cite{neda2003flatness}.

In the following, we will use numerical examples to illustrate the curve shift associated with different errors. In all the figures, the curves that represent the intersections of the true FDOA cones with the WGS84 ellipsooid are marked in yellow and the curves calculated with respect to the errors are marked in red.

\subsection{The UAV cases}
According to the U.S. Department of Defense \cite{dempsey2010eyes}, UAVs can be conveniently classified into five categories according to their sizes (see Appendix for the classifications) . In our application scenarios, a medium- or large-size UAV is capable of carrying on the tasks. In this subsection, we assume that the UAV speed and height are 50$\text{m/s}$ (180$\text{km/h}$) and 2000$\text{m}$, respectively. The yaw angle of the UAV is selected as \ang{0}, the pitch angle is selected as \ang{-30} and the roll angle of the UAV is selected as \ang{0}. We consider three situations associated with different magnitudes (close to \ang{30}, close to \ang{0} and close to \ang{90}) of the semi-angle of the true FDOA cone, to better describe the situations in practical scenarios.

\subsubsection{Curve shift associated with the error between the nominal frequency and the true frequency}
\begin{itemize}
\item \textbf{UAV example 1:}
Assume that the measured frequency at the receiver is 299,792,501.33$\text{Hz}$, and the true frequency of the emitter is 299,792,468$\text{Hz}$. There is a 10Hz error between the nominal frequency and the true frequency of the emitter. The semi-angle of the true FDOA cone is calculated according to the true frequency of the emitter, which can be obtained as \ang{48.243}. Since the receiver has only the knowledge of the nominal frequency of the emitter, the semi-angle of the FDOA cone calculated at the receiver is \ang{29.934}. We plot the two intersection curves in Google Earth, see Fig. \ref{fig:E1}, and observe that the magnitude of the curve shift ranges from approximately 730m to 4,000km \footnote{Note that we do not consider the fact that the signal from the emitter can be blocked due to the eccentricity of the earth surface. The curve shift does not represent the error of the position of the emitter.}.

\begin{figure}[t]
\centering
\includegraphics[width=0.9\linewidth]{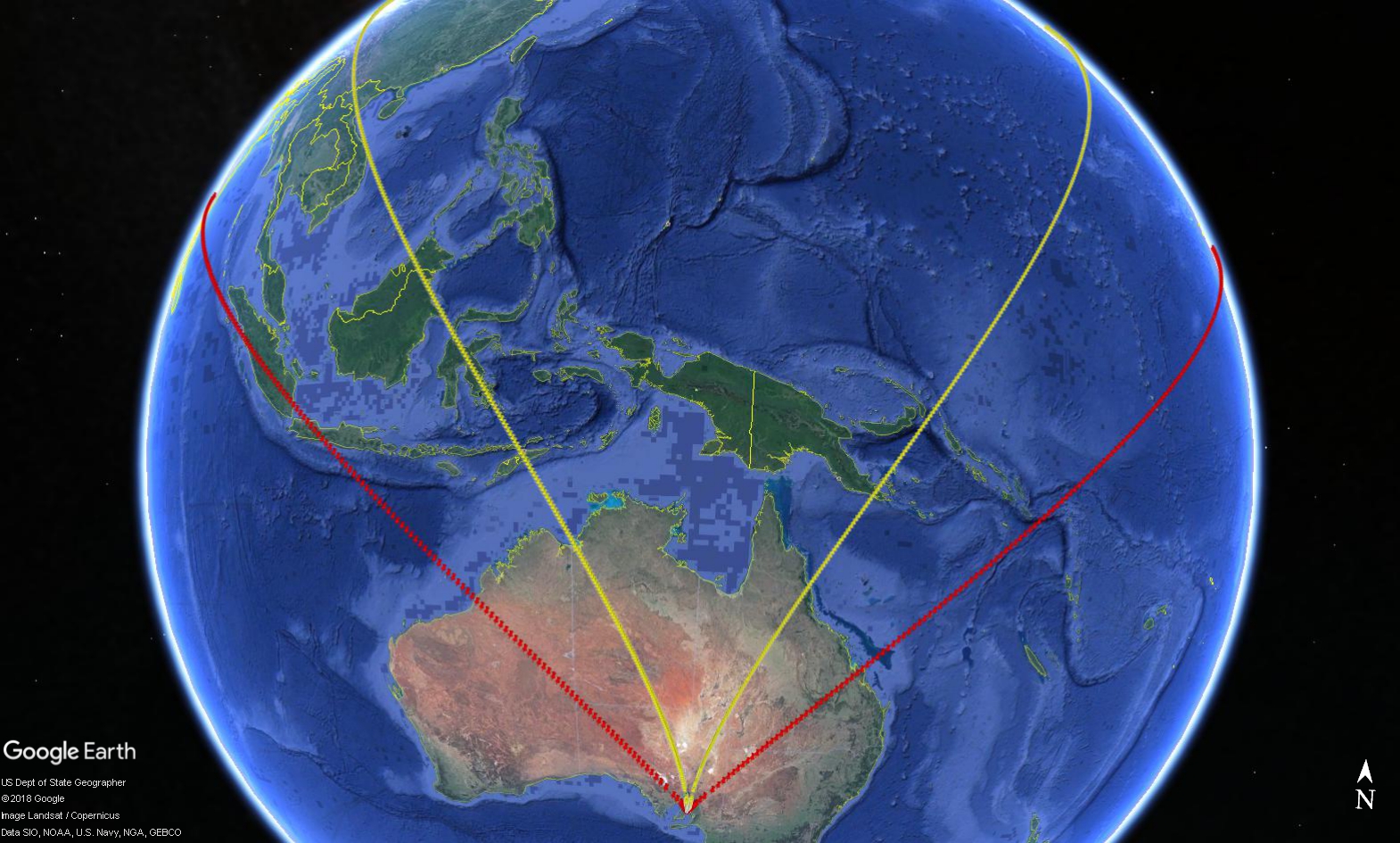}
\caption{\label{fig:E1}UAV example 1.}
\end{figure}

\item \textbf{UAV example 2:}

We still assume that the true frequency of the emitter is 299,792,468$\text{Hz}$. By assuming that the measured frequency at the receiver is 299,792,507.93Hz, the Doppler shift calculated according to the nominal frequency and the true frequency of the emitter are 49.93Hz and 39.93Hz, respectively. The semi-angle of the FDOA cone associated with the true Doppler shift is \ang{3}. The semi-angle calculated according to the difference between the measured frequency and the nominal frequency is \ang{37.0}. The two intersection curves for this example are depicted in Fig. \ref{fig:E2}. The minimum magnitude of the curve shift is approximately 2,250m and the maximum magnitude of the curve shift is too large to be determined in Google Earth. This is because the size of the yellow curve is very small (the radius is about 800m) and the red curve in Fig. \ref{fig:E2} has a similar shape as the red curve in Fig. \ref{fig:E1}, with a much bigger size. 

\begin{figure}[t]
\centering
\includegraphics[width=0.9\linewidth]{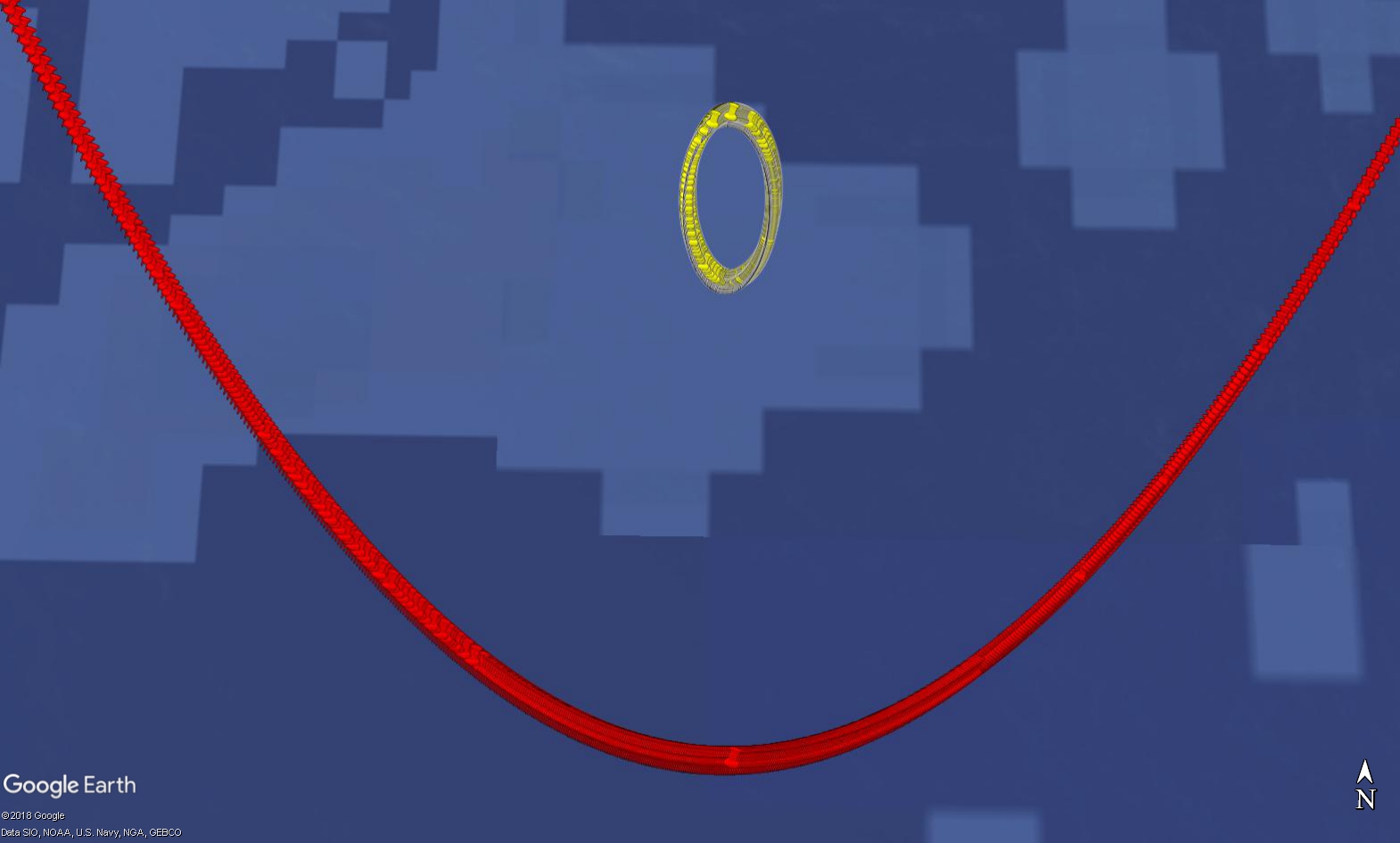}
\caption{\label{fig:E2}UAV example 2. }
\end{figure}

\item \textbf{UAV example 3:}

Assume that the measured frequency at the receiver is 299,792,470.615Hz, and the true frequency of the emitter is 299,792,468Hz. The true Doppler shift is 2.615Hz, thus the semi-angle of the true FDOA cone is \ang{87}. The Doppler shift calculated according to the nominal frequency of the emitter is 12.615 Hz and its associated FDOA cone has a semi-angle with the value of \ang{75.386}. The magnitude of the curve shift measured in Goolge Earth ranges from approximately 500m to 2,600km
\begin{figure}[t]
\centering
\includegraphics[width=0.9\linewidth]{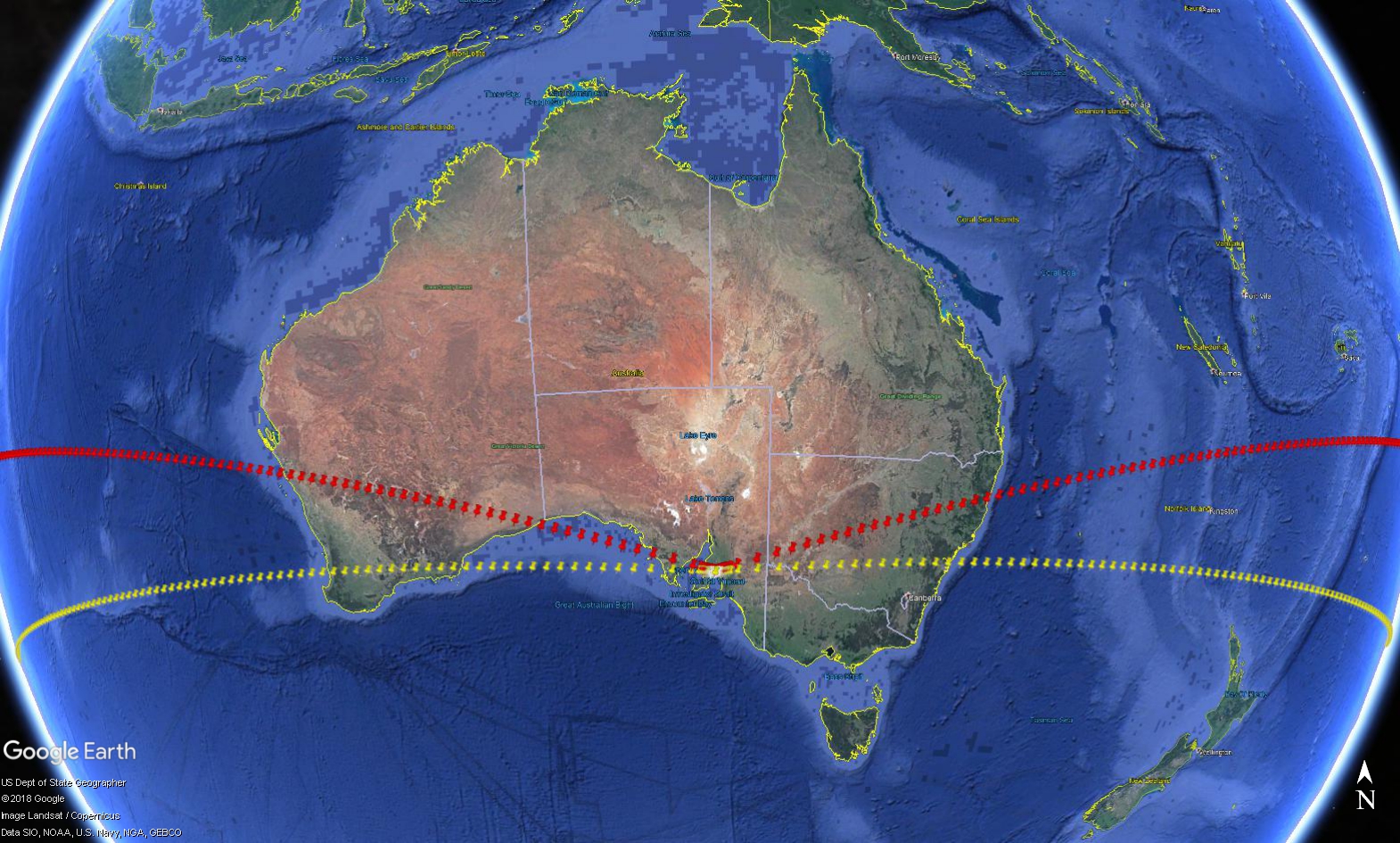}
\caption{\label{fig:E3}UAV example 3.}
\end{figure}

\end{itemize}

By observing the above three UAV examples, we can conclude that there is a significant curve shift associated with the error between the nominal frequency and the true frequency of the emitter. When applying our algorithm on the UAV platforms, one must be very careful to ensure that the knowledge of the transmission frequency of the emitter is accurate enough.  

\subsubsection{Curve shift associated with the changing refractive index of the atmosphere}

The UAV operation altitude in our application scenarios is less than 5km. According to the air refractive model proposed in \cite{neda2003flatness}, the air refractive index can be regarded as a constant with a value 1.0003. We will use this value in the following examples.  

\begin{itemize}
\item \textbf{UAV example 4:}

Assume that the Doppler shift measured at the receiver is 43.3Hz. Then the semi-angle of the FDOA cone calculated using the speed of electromagnetic wave in the vacuum is \ang{30}. By considering the refractive index of the air, we can directly substitute \eqref{eq:light_speed_in_air} into equation \eqref{eq:cone_semi_angle} and obtain that the semi-angle of the FDOA cone considering the refractive index of the air is \ang{29.973}. We depeict the two intersections curves in Fig. \ref{fig:E4}. The range of the curve shift varies from 3m to 6,500m.   

\begin{figure}[t]
\centering
\includegraphics[width=0.9\linewidth]{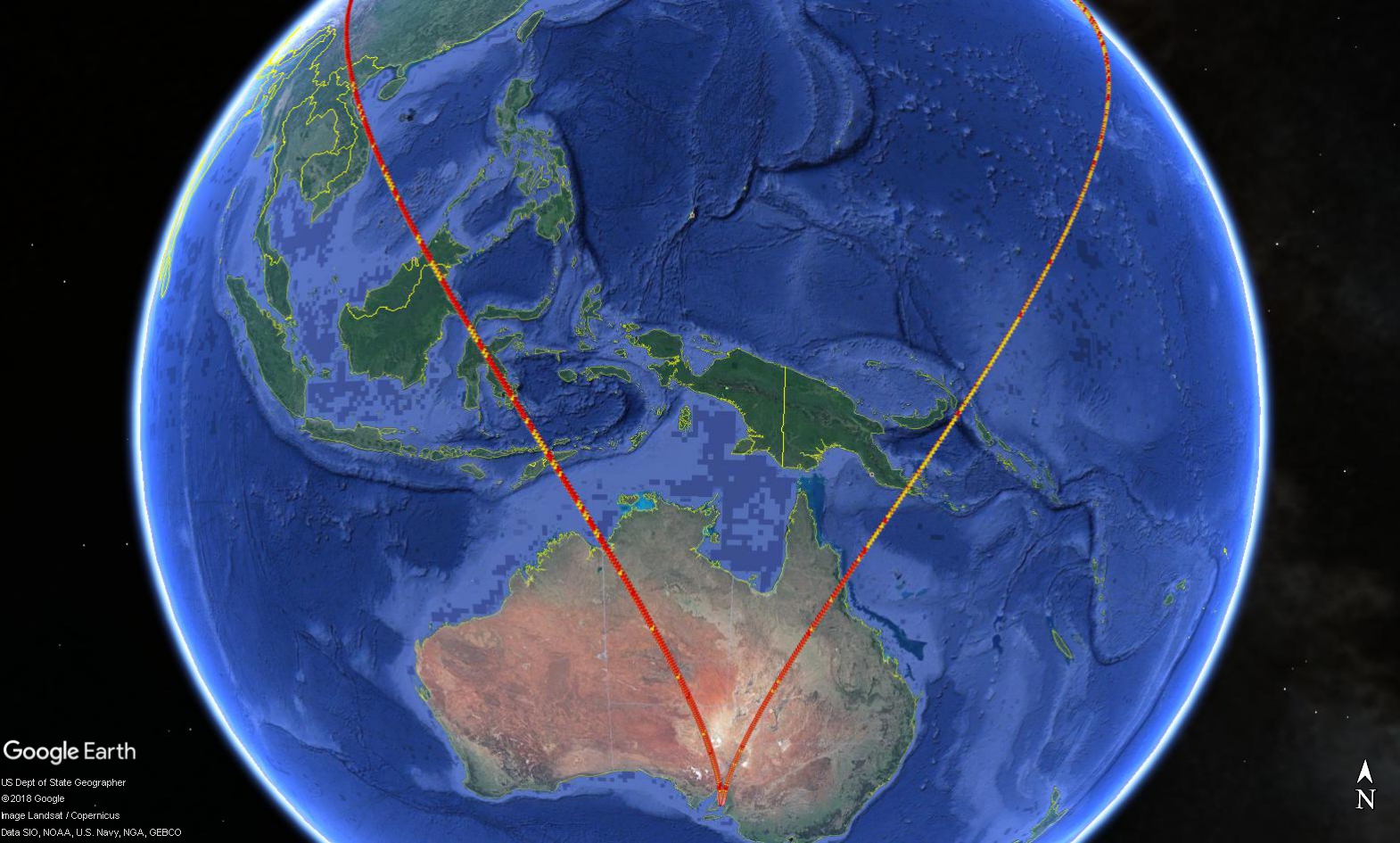}
\caption{\label{fig:E4}UAV example 4}
\end{figure}

\item \textbf{UAV example 5:} 

In this example, we assume that the Doppler shift measured at the receiver is 49.93Hz. Thus the semi-angle of the FDOA cone calculated using the speed of the electromagnetic wave in the vacuum is \ang{3}. Considering the speed of the electromagnetic wave in the air, the new magnitude of the semi-angle of the FDOA cone can be obtained as \ang{2.688}. Fig. \ref{fig:E5} shows the two intersection curves. The curve shift of the two curves varies from 20m to 40m.   

\begin{figure}[t]
\centering
\includegraphics[width=0.9\linewidth]{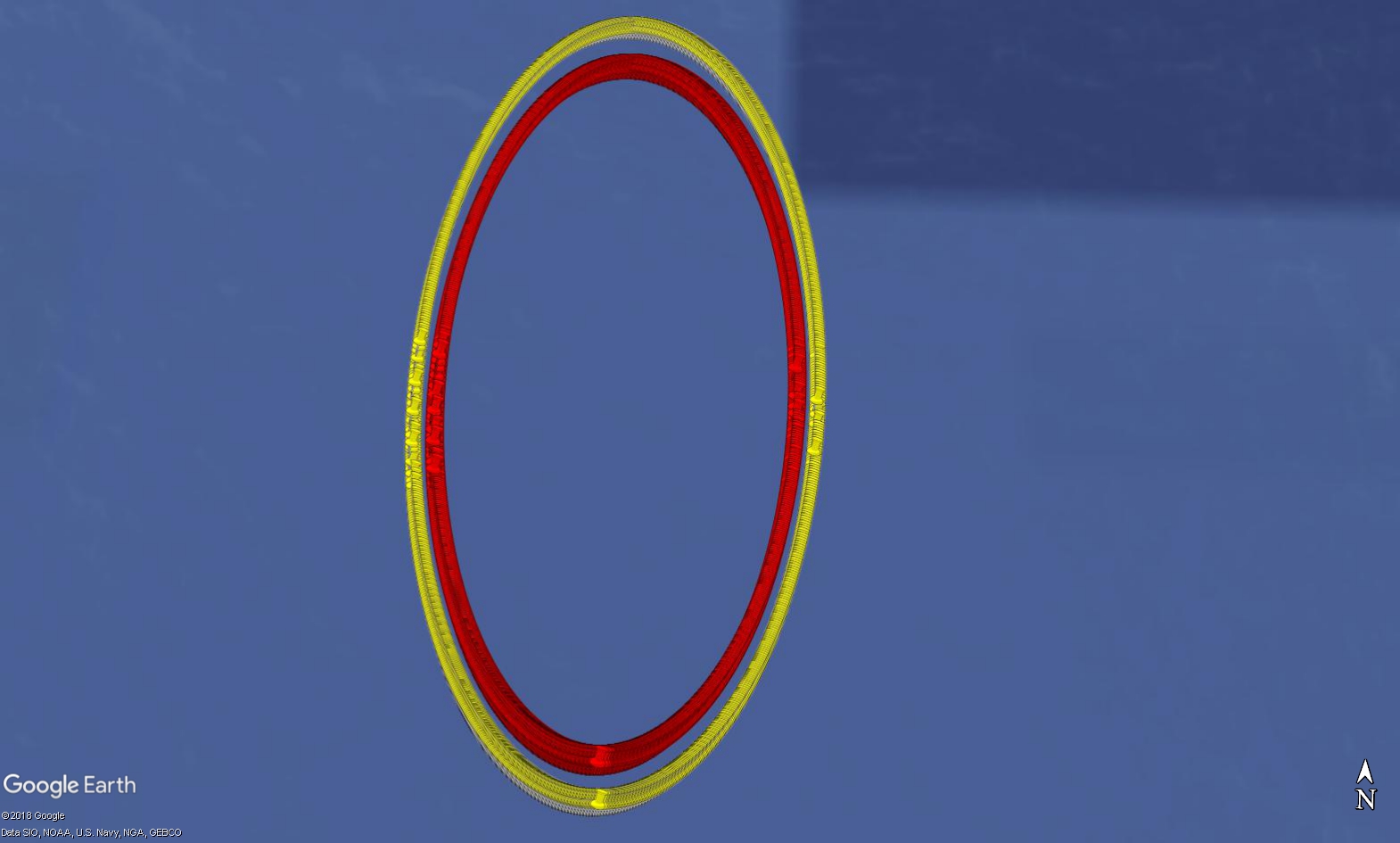}
\caption{\label{fig:E5}UAV example 5}
\end{figure}

\item \textbf{UAV example 6:} 

In this example, we assume that the Doppler shift measured at the receiver is 2.615Hz. The semi-angle of the FDOA cone calculated using the speed of an electromagnetic wave in a vacuum is \ang{87}. Considering the speed of an electromagnetic wave in the air, the magnitude of the semi-angle of the true FDOA cone is \ang{86.993}. The curve shift of the two curves varies from 3m to 1,500m.  

\begin{figure}[t]
\centering
\includegraphics[width=0.9\linewidth]{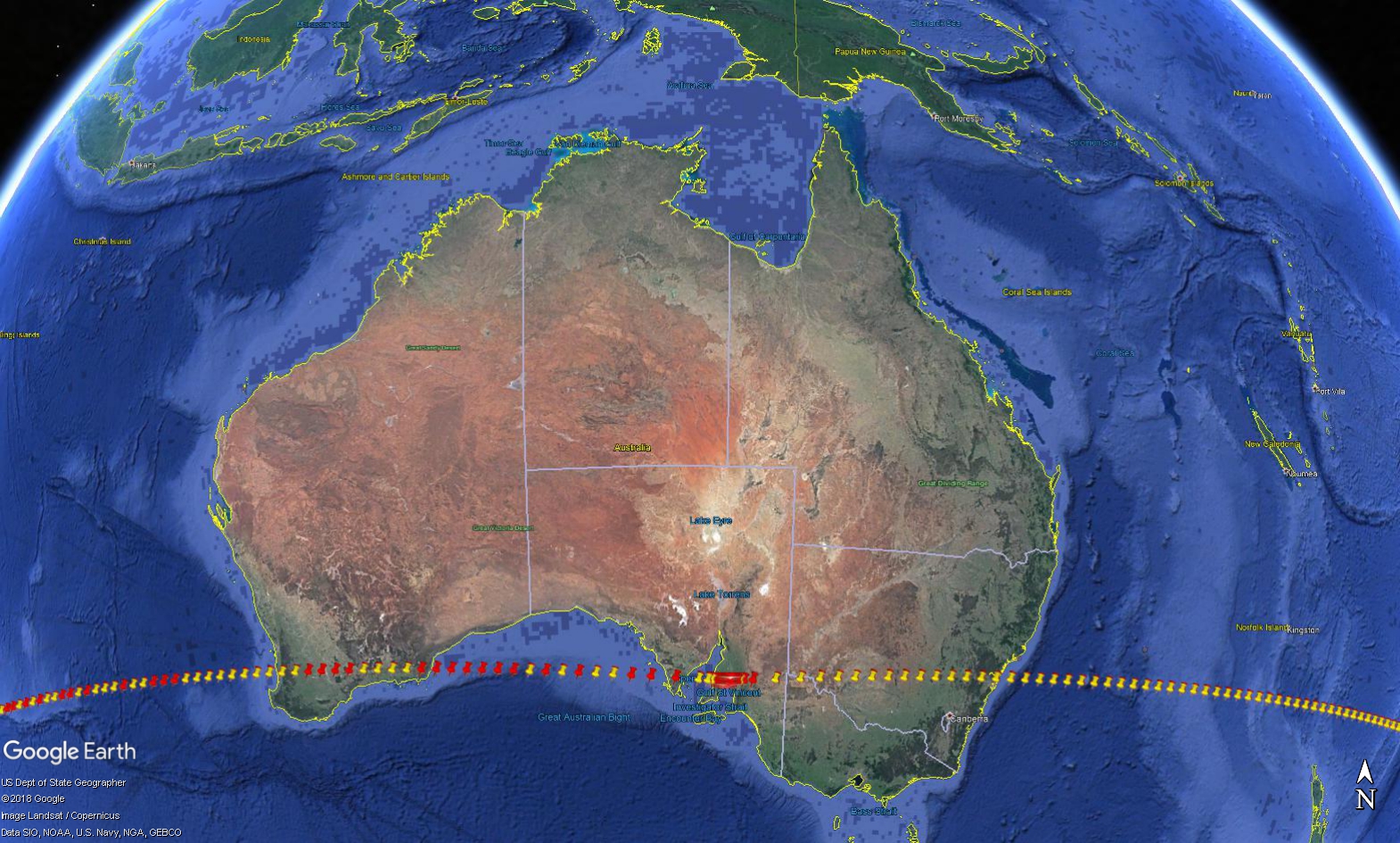}
\caption{\label{fig:E6}UAV example 6}
\end{figure}

\end{itemize}

{According to the three examples for UAV cases considering the changing refractive index of the atmosphere, we can find that the magnitude of the curve shift is actually minor. Though the maximum curve shifts in example 4 and 6 are 1500m and 3500m, they both appear at the positions far away from the receiver (the distance can be more than 4,000km). And it is obvious that signal emitted at those positions cannot be received by the receiver. In conclusion, we may neglect the effect of the changing refractive index of the atmosphere when applying our algorithm in practice.}

\subsection{The LEOS cases}

\subsubsection{Curve shift associated with the error between the nominal frequency and the true frequency of the emitter}
In all the examples for LEOS, we assume that the LEOS is above the equator and the roll, pitch and yaw are all selected as \ang{0}. We also assume that the satellite speed is 7800m/s (28,080km/h). The height of the satellite is 200,000m (200km). These two parameters are typical for LEOS. In all the examples provided below, we assume that the error between the nominal frequency and the true frequency is 60Hz.

\begin{itemize}
\item\textbf{LEOS example 1:}
The true transmission frequency of the emitter is 299,792,518Hz. Suppose that the measured frequency at LEOS is 299,798,033.4329Hz. Thus the true Doppler shift is 5515.4329Hz. The semi-angle of the FDOA cone calculated according to the true frequency of the emitter is \ang{45}. The semi-angle of the FDOA cone calculated according to the nominal frequency of the emitter $45.62^o$. The two curves are shown in Fig. \ref{fig:LE1}. The curve shift of the two curves varies from approximately 5,000m to 15,000m.

\begin{figure}[tb]
\centering
\includegraphics[width=0.9\linewidth]{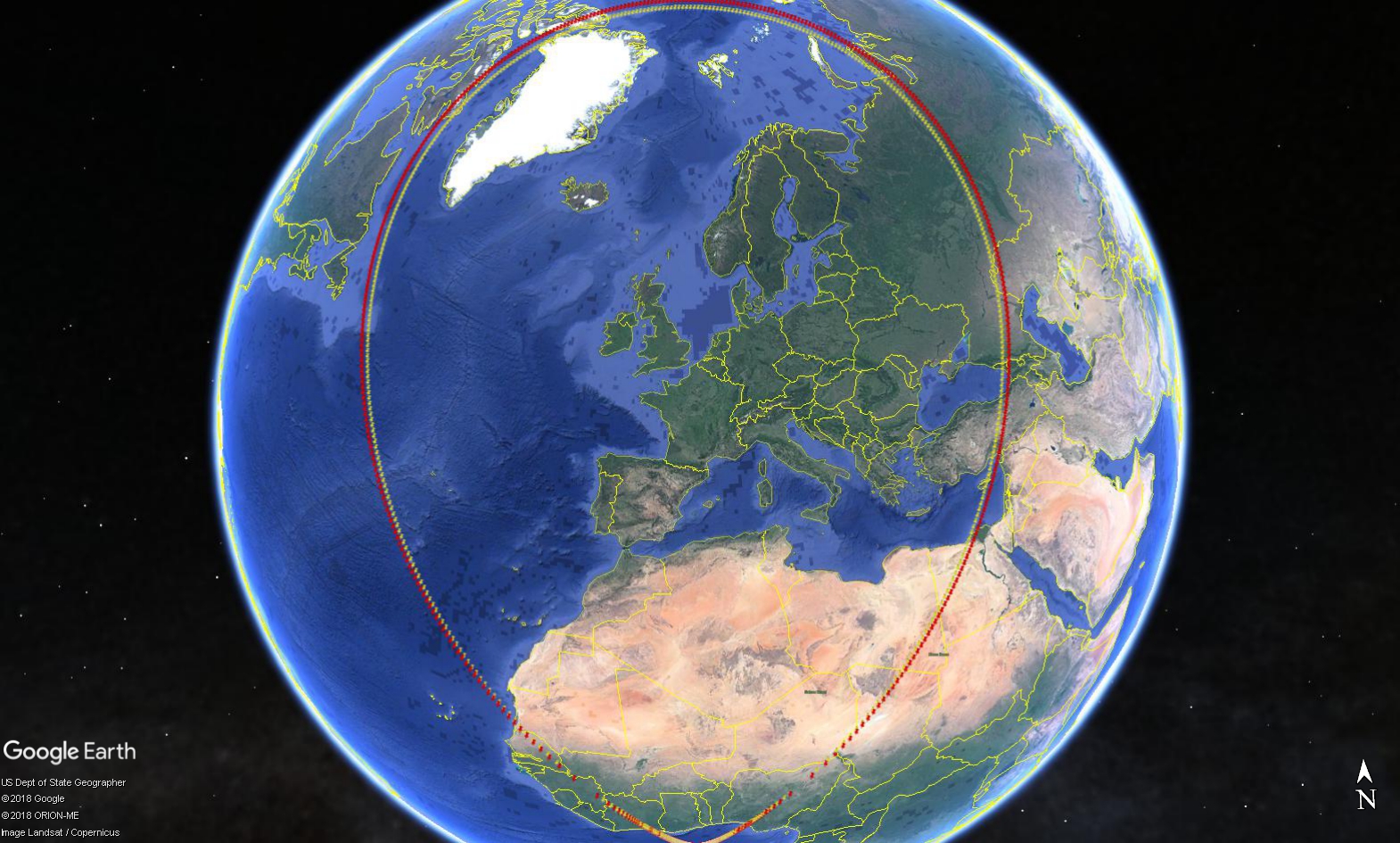}
\caption{\label{fig:LE1}LEOS example 1}
\end{figure}

\item\textbf{LEOS example 2:} 

Since the LEOS orbit is very close to circular and the moving direction of LEOS is close to be parallel to the plane of the horizon. If we consider the situation that the semi-angle of the FDOA cone is close to \ang{0}, then it is impossible to find an emitter on the earth's surface to generate such the FDOA cone. We do not need to provide illustrations to this example. 

\item\textbf{LEOS example 3:}
We still assume that the true transmission frequency of the emitter is 299,792,518Hz. Assume that the Doppler shift measured at the LEOS is 347.94Hz. Because of the error between the nominal frequency and the true frequency, the output of the Doppler shift at the LEOS is 407.94Hz. Then the semi-angle of the FDOA cone calculated according to the nominal frequency of the emitter is $\ang{87.433}$. The semi-angle of the FDOA cone calculated according to the true transmission frequency of the emitter is \ang{87}. The two curves are depicted in Fig. \ref{fig:E6}. The curve shift of the two curves varies from 2,000m to 100,000m.

\begin{figure}[tb]
\centering
\includegraphics[width=0.9\linewidth]{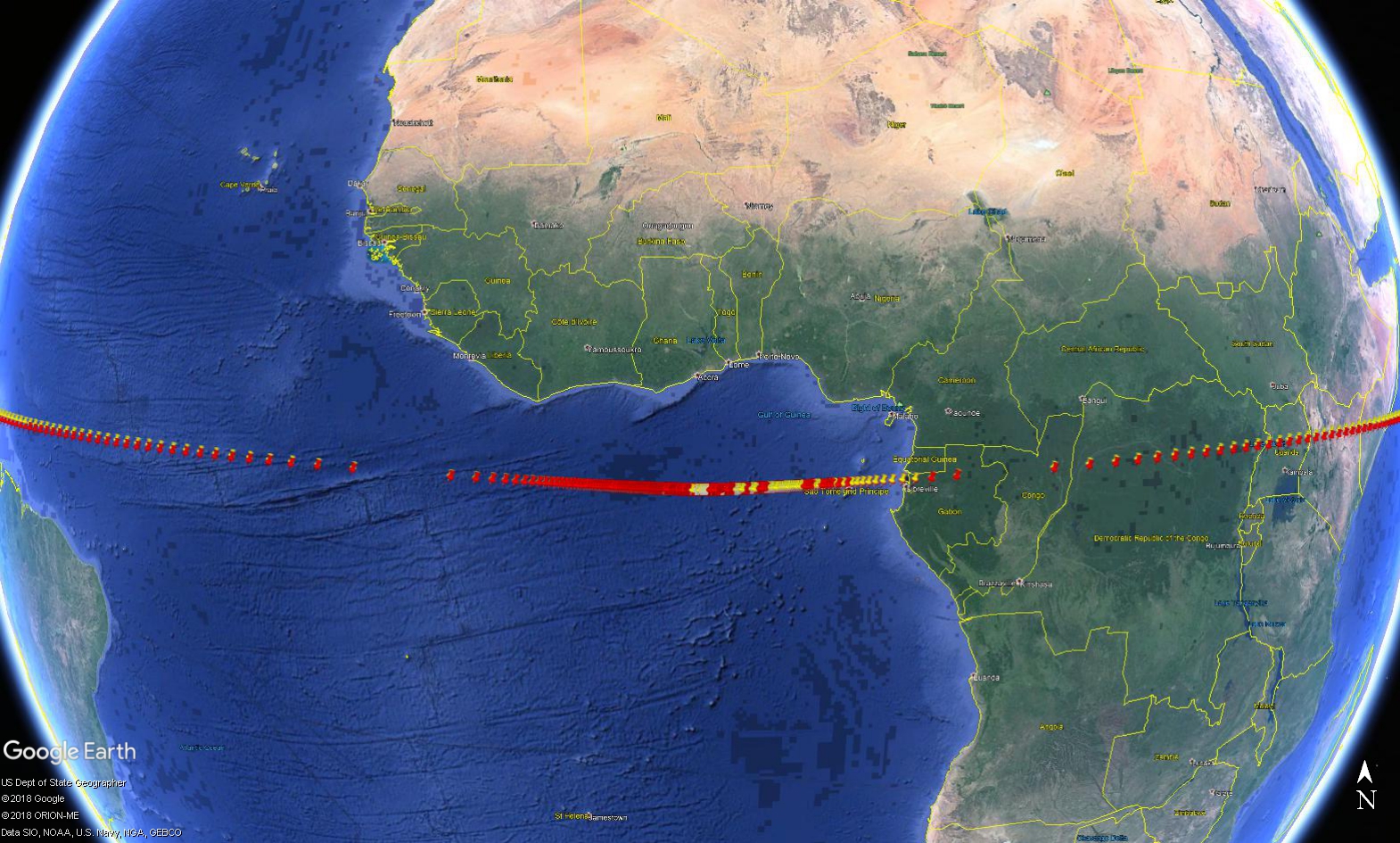}
\caption{\label{fig:LE3}LEOS exmaple 3}
\end{figure}

\end{itemize}

By observing the above LEOS examples, we can find out that the magnitude of the curve shift is not negligible, though the change of the semi-angle of the FDOA cone is minor. This is due to the fact that LEOS are operated at a high altitude. When applying our algorithms on the LEOS, it is still required to have a accurate knowledge of the true transmission frequency of the emitter.

\subsubsection{Curve shift associated with the changing refractive index of the atmosphere}
The refractive index of air varies according to a function of temperature and pressure, which leads to a refractive index gradient in the atmosphere. Although this gradient is very small, atmospheric refraction of electromagnetic waves is observable due to the large distances traveled in the atmosphere between the transmitter and the emitter. 

The earth atmosphere can be divided into several layers, including Troposphere, Stratosphere, Mesosphere and Ionosphere. From the viewpoint of atmospheric refraction, only the first two layers, the troposphere and the stratosphere, are important. In the upper layers of the atmosphere the air is so rarefied that the refractive index can be considered to be unity.

In this subsection, we roughly estimate the curve shift by assuming a simple two layer atmosphere model instead of the complicated model described in \cite{neda2003flatness}. Our model is described as follows: assume that the height of the Troposphere is $20km$, and the Stratosphere starts at $20km$ and ends at $50km$. The refractive index is assumed to be a constant for each layer, i.e, the refractive index $n_s$ of the Stratosphere is equal to 1 and the refractive index $n_t$ of Troposphere is equal to 1.0003. 

\begin{figure}[tb]
\centering
\includegraphics[width=0.9\linewidth]{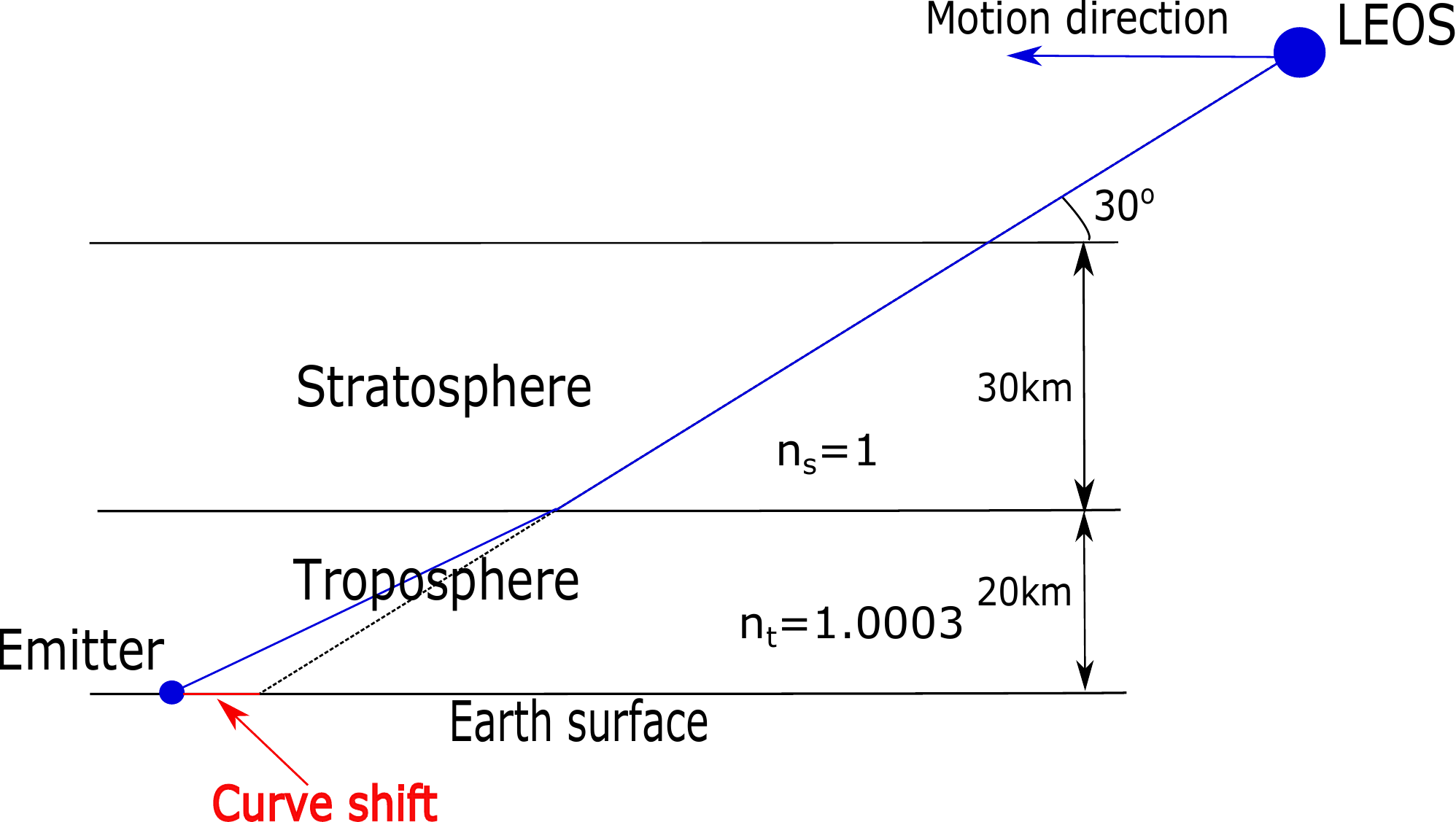}
\caption{\label{fig:f10}Graphical illustration of the assumed scenario of the atmosphere.}
\end{figure}

Now we describe our scenario to calculate an estimated curve shift. The graphical illustration of this scenario is in Fig. \ref{fig:f10}. Suppose the LEOS is moving horizontally and its calculated semi-angle of the FDOA cone associated with one single measurement is \ang{30}. According to the above 2-layer atmosphere model and the well-known Snell's law, it is straightforward to obtain that the curve shift is about 41m, which is small when compared to the curve shift associated with the error between the nominal and the true frequencies.

\subsubsection{Curve shift associated with the relativistic Doppler effect}
The relativistic Doppler effect is the change in frequency of electromagnetic waves, caused by the relative motion of the source and the observer (as in the classical Doppler effect), when taking into account effects described by the special theory of relativity. The relativistic Doppler effect is different from the non-relativistic Doppler effect as the equations include the time dilation effect of special relativity and do not involve the medium of propagation as a reference point. 

Due to relativistic effects, clocks on the receiver are time dilated relative to clocks at the source. The measured frequency in the receiver, denoted by $f_{r}$, can be described by 
\begin{equation}
f_{r} = \rho f 
\label{eq:relativistic_doppler}
\end{equation}
where $f$ is the classical Doppler shift described by \eqref{eq:generalDoppler}. $\rho$ is called the Lorentz factor, which is defined as 
\begin{equation}
\rho = \frac{1}{\sqrt{1-\varsigma^2}}
\end{equation}
in which $\varsigma:=v_r/c$ is the velocity of the receiver in terms of the speed of light.

We now use an example to illustrate the curve shift associated with the relativistic Doppler shift. We use the parameters selected in LEOS example 1 in this example. We assume that there exists a FODA cone with an accurate semi-angle of \ang{45} associated with a single Doppler shift measurement at the LEO satellite. Now we consider the relativistic Doppler effect. According to \eqref{eq:relativistic_doppler}, it can be obtained that $\rho$ is equal to a value that is very close to 1 with the error less than $10^{-10}$. By using this value, we can obtain that the change of the semi-angle of FDOA cone is $1.94\times 10^{-7}$ degrees. By assuming that the earth surface is flat and there is no refraction through the air, it can be directly calculated that the curve shift is about $0.0001352m$, which is small enough to be neglected.

 
 

\section{Conclusions and future work}
This paper investigates the problem of determining a curve on the earth's surface on which a stationary emitter must lie based only on the measured frequency on a mobile vehicle (UAV or LEOS) and the vehicle's navigation data (position and velocity). Our investigated scenario considers the following two facts: 1) Coriolis effect; 2) the bumpy earth's surface. We prove that the relative velocity component due to the Coriolis effect makes zero contribution to the Doppler shift. We then provide methods for building equations to describe a FDOA cone and finding its intersection curve with the WGS84 ellipsoid. The curve comprises points, which are mapped into a DTED dataset to find the curve on the earth's terrain. We further provide numerical examples to illustrate how the errors resulting from the nonconstant refractive index of the atmosphere and from lack of precise knowledge of the transmitter frequency affect the positions of curves. Future work will focus on estimating the true frequency of the emitter using multiple frequency measurements and their associated intersection curves on the surface of the WGS84 ellipsoid.

\bibliographystyle{ieeetr}
\bibliography{WGS.bib}

\clearpage
\appendix
\subsection{Abbreviations}
\begin{tabularx}{\textwidth}{XX}\hline
Abbreviation & Word/phrase\\ \hline
ECEF & Earth-Centered, Earth-Fixed\\
ENU & East North Up \\
FDOA & Frequency Difference of Arrival\\
GPS & Global Positioning System \\ 
INS & Inertial Navigation System\\ 
LEOS & Low Earth Orbiting Satellite\\
MSL & Mean Sea Level  \\
WGS & World Geodetic System\\ 
TDOA & Time Difference of Arrival\\
UAV & Unmanned Aerial Vehicle\\
\hline
\end{tabularx}

\subsection{Notations}
\begin{tabularx}{\textwidth}{XXX}\hline
Notation & Description & Magnitude\\ \hline
$a$ & semi-major axis of WGS84 ellipsoid & 6378137.0m\\
$b$ & semi-minor axis of WGS84 ellipsoid & 6356752.314245m\\
$c$ & speed of the electromagnetic wave in the vacuum & $299792458\text{m/s}$\\
$d$ & parameter of an arbitrary right circular cone & N/A\\
$e^2$ & the square of first eccentricity & $6.69437999014\times 10^{-3}$\\
$f$ & flattening of the reference ellipsoid & 1/298.257223563\\
$f_0$ & unshifted transmitter frequency & N/A\\
$f_r$ & received frequency & N/A\\
$h$ & ellipsoid height& N/A\\
$n$ & refractive index of air & N/A\\
$H$ & orthometric height& N/A\\
$N$ & geoid height& NA\\
$\mathbf{e}^v$ & unit vector in the direction of velocity in ECEF & N/A\\
$\mathbf{p}$ & emitter position in ECEF & N/A\\
$\mathbf{r}$ & receiver position in ECEF & N/A\\
$\mathbf{v}$ & vehicle velocity & N/A\\
$\alpha$ & roll angle &  N/A\\
$\beta$  & pitch angle &  N/A\\
$\gamma$ & yaw angle & N/A\\
$\phi$ & latitude &  N/A\\
$\lambda$  & longitude &  N/A\\
$\rho$ & Lorentz factor & N/A\\
$\mathbf{\Omega}$ & angular velocity & N/A\\
\hline
\end{tabularx}
\\

\subsection{UAVs Classification}
\begin{tabularx}{\textwidth}{XXXX}\hline
Size & Max-takeoff Weight (lbs) & Operating Altitude (ft) & Airspeed (knots)\\ \hline
Small   &    0-20               & $< 1200$                   & $< 100$                       \\ 
Medium  &    21-55              & $< 3500$                   & $< 250$                       \\ 
Large   &    $< 1320$           & $< 18000$                  & $< 250$                       \\ 
Larger  &    $> 1320$           & $< 18000$                  & Any airspeed               \\ 
Largest &    $> 1320$           & $> 18000$                  & Any airspeed               \\
\hline
\end{tabularx}



\end{document}